\documentclass[pdflatex,sn-mathphys-num]{sn-jnl}

\usepackage{graphicx}%
\usepackage{multirow}%
\usepackage{amsmath,amssymb,amsfonts}%
\usepackage{amsthm}%
\usepackage{mathrsfs}%
\usepackage[title]{appendix}%
\usepackage{xcolor}%
\usepackage{textcomp}%
\usepackage{manyfoot}%
\usepackage{booktabs}%
\usepackage{algorithm}%
\usepackage{algorithmicx}%
\usepackage{algpseudocode}%
\usepackage{listings}%
\usepackage{graphicx}
\usepackage{subcaption}

\theoremstyle{thmstyleone}%
\theoremstyle{thmstyletwo}%

\theoremstyle{thmstylethree}%

\usepackage[font=small,labelfont=bf]{caption}
\newcommand{\affilsmall}{\footnotesize}

\begin{document}

\title[Singular geometry and eigenframe topology in local rank-2 tensor observables]{Singular geometry and eigenframe topology in local rank-2 tensor observables}


\author*[1,4]{\fnm{I. C. J.} \sur{Yap}}\email{ian.chang.jie.yap@uni-due.de}
\equalcont{{\footnotesize These authors contributed equally to this work.}}

\author[2,3]{\fnm{S. Q.} \sur{Jin}}
\equalcont{{\footnotesize These authors contributed equally to this work.}}

\author[1]{\fnm{B.} \sur{Dörschel}}
\equalcont{{\footnotesize These authors contributed equally to this work.}}

\author[3]{\fnm{T. T.} \sur{Dang}}

\author[5]{\fnm{P. M.} \sur{Scott}}

\author[4]{\fnm{H. C.} \sur{Hofsäss}}
\equalsup{{\footnotesize These authors supervised this work equally.}}

\author[1]{\fnm{D. C.} \sur{Lupascu}}
\equalsup{{\footnotesize These authors supervised this work equally.}}

\author[3]{\fnm{A.} \sur{Krawczuk}}
\equalsup{{\footnotesize These authors supervised this work equally.}}

\author[1,5]{\fnm{J. H.} \sur{Schell}}
\equalsup{{\footnotesize These authors supervised this work equally.}}

\affil[1]{{\affilsmall
\orgdiv{Institute of Materials Science and Center for Nanointegration Duisburg--Essen (CENIDE)},
\orgname{University of Duisburg--Essen},
\orgaddress{\city{Essen}, \country{Germany}}}}

\affil[2]{{\affilsmall
\orgname{Max Planck Institute for Multidisciplinary Sciences},
\orgaddress{\city{Göttingen}, \country{Germany}}}}

\affil[3]{{\affilsmall
\orgdiv{Institute of Inorganic Chemistry},
\orgname{Georg-August University of Göttingen},
\orgaddress{\city{Göttingen}, \country{Germany}}}}

\affil[4]{{\affilsmall
\orgdiv{II. Physics Institute},
\orgname{Georg-August University of Göttingen},
\orgaddress{\city{Göttingen}, \country{Germany}}}}

\affil[5]{{\affilsmall
\orgname{European Organization for Nuclear Research (CERN)},
\orgaddress{\city{Geneva}, \country{Switzerland}}}}


\abstract{\footnotesize\unboldmath Many observables are symmetric second-rank tensors, reported through magnitude-ordered principal values and axes. This chart folds tensor space: the parameters develop cusps and exchange labels where the tensor is smooth. It also hides a global effect: an arrow carried along a principal axis around a loop encircling a degeneracy can return reversed, defining a binary return parity, invariant under smooth deformations of the loop avoiding degeneracy. For tensor fields, this structure is long established, but the parameters are coordinates of the domain on which the field is defined, and the degeneracies are a feature of that particular field. Here we show that the electric field gradient (EFG) carries the same structure in a control space: a traceless observable at a single probe site, steered through its five-dimensional tensor space by symmetry-adapted strain, with the encircling loop applied rather than found. First-principles calculations reveal an isolated degeneracy with nontrivial parity in rutile TiO$_2$, point- or line-like degeneracies in SnO$_2$ depending on the control slice, and access to all five EFG components in cubic MgO. Thus, return parity becomes accessible for a local observable, and strain-tuned, orientation-resolved hyperfine spectroscopy offers a route to reconstruct it.}




\maketitle
\begin{figure}[H]
    \centering
    \includegraphics[width=0.8\linewidth]{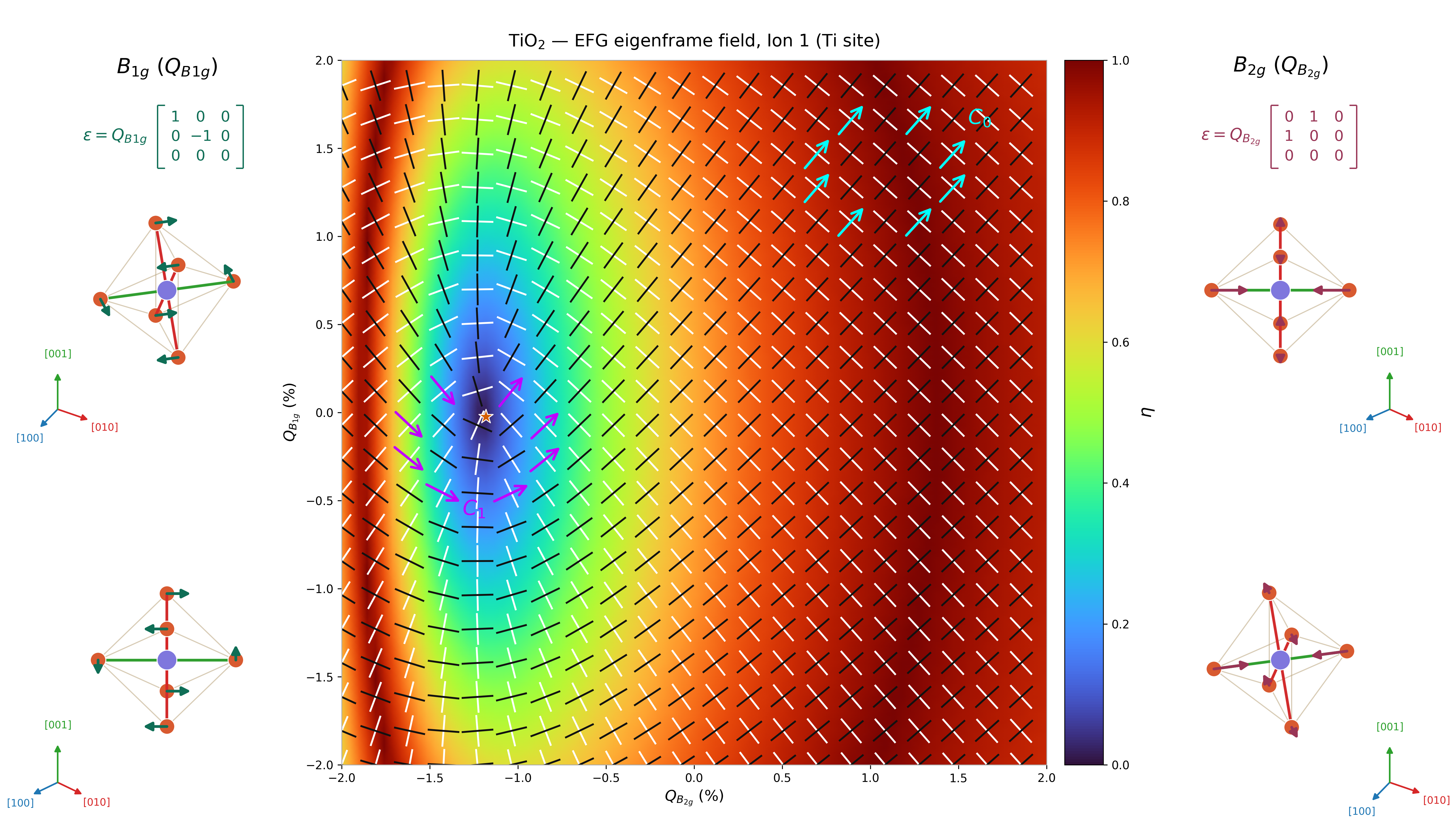}
    \caption{
    {\bfseries\boldmath Sign reversal of a transported principal axis of the EFG tensor at the Ti site of rutile TiO$_2$, in the $Q_{B_{2g}}$--$Q_{B_{1g}}$ strain-control plane.}
    The colour map shows the asymmetry parameter $\eta$; the black and white line segments show the two in-plane principal axes of the EFG tensor. Each principal axis is an axis without an arrow: $\overrightarrow{v}$ and $-\overrightarrow{v}$ are physically indistinguishable. The auxiliary arrows show one such axis after an arrow has been chosen and carried continuously around each of the two closed loops, labelled $C_0$ and $C_1$. We record the outcome of each loop by its return parity $w_1$: $w_1=0$ if the arrow returns as it started, and $w_1=1$ if it returns reversed. The loop $C_0$ does not enclose the degeneracy point (star; there $\eta=0$, two principal values coincide, and the axes become undefined); the arrow returns as it started, giving $w_1(C_0)=0$. The loop $C_1$ encloses the degeneracy point, and the arrow returns reversed, giving $w_1(C_1)=1$. This parity is a binary topological property of the loop called the first Stiefel--Whitney invariant, and cannot be changed by any smooth choice of arrows. The $\eta=1$ ridges are ordering seams $\Delta_0$ of the reporting convention; they cause local relabellings but no sign reversal. The shown control plane is created by the two symmetry-adapted strain modes $Q_{B_{1g}}$ and $Q_{B_{2g}}$. The corresponding strain tensors and their effects on the local TiO$_2$ environment are illustrated on the left and right of the colour map, respectively. For each strain mode, the distorted environment is shown from two different projections to make the atomic displacements more apparent. The central titanium atom is shown in purple and the surrounding oxygen atoms in orange. Green lines connect the titanium atom to the two oxygen atoms in the horizontal plane, while red lines connect it to the four oxygen atoms associated with the vertical coordination directions. The crystallographic directions $[100]$, $[010]$, and $[001]$ are indicated for each projection.
    }
    \label{fig:holonomy}
\end{figure}

\newpage

\section{Spectral topology of local tensor observables}\label{sec1}
Symmetric second-rank tensors are among the most common observables in physics. Stress and strain in materials~\cite{Schroeder_J_2024}, diffusion tensors in imaging~\cite{Basser_P_J_1994}, magnetic-susceptibility~\cite{Cho_H_2017} and electrical-conductivity~\cite{Yan_H_2020} tensors in anisotropic media, the gyration tensor of optical activity in gyrotropic crystals~\cite{Nye_J_F_1985}, the symmetric part of distributed atomic polarizabilities in quantum crystallography~\cite{Krawczuk_A_2014}, and the hyperfine tensors of Nuclear Magnetic Resonance (NMR)~\cite{Jerschow_A_2005}, Electron Paramagnetic Resonance (EPR)~\cite{Abragam_Bleaney_1970}, M\"ossbauer~\cite{Liu_Y_S_1984}, and time-differential perturbed angular correlation (TDPAC)~\cite{Schell_J_2017,Kaufmann_E_N_1979} spectroscopy are all real symmetric $3\times3$ tensors. Most carry a nonzero trace, but subtracting the isotropic part leaves principal axes and eigenvalue degeneracies unchanged, so everything that follows depends only on the deviatoric part~\cite{Nye_J_F_1985}. 

In practice, such a tensor is almost never reported in full. It is diagonalized, its principal values are sorted by a convention, and the result is compressed into a small set of reported parameters such as the largest principal component $|V_{33}|$ and the asymmetry parameter $\eta$~\cite{Kaufmann_E_N_1979,Harris_R_K_2008,Nardelli_F_2020}. We refer to any such reduced description as a chart: a coordinate system on the space of tensors, convenient for reporting but not guaranteed to be faithful everywhere. Diagonalization creates a representation with a geometry of its own.

The chart becomes problematic when the tensor is tracked continuously under temperature, strain or composition: smooth physical evolution can appear as sharp dips, cusps or label exchanges \cite{Harris_R_K_2008,Derevianko_A_2026,Kanert_0_1988,PRB_paper}. Two distinct loci organize these artifacts. The first is the degeneracy stratum $\Sigma_0$, the subset of tensor space where two principal values are equal. There, the corresponding principal axes are no longer unique and $\eta=0$. The second is the determinant-zero set $\Delta_0$, where one principal value vanishes. Under magnitude ordering, $\Delta_0$ becomes an ordering seam: the reporting convention switches labels and $\eta=1$, even though the underlying tensor remains smooth and nondegenerate.

A subtler phenomenon appears for closed paths. Each principal axis of a symmetric tensor is an axis without an arrow: $\overrightarrow{v}$ and $-\overrightarrow{v}$ are physically indistinguishable, because each axis enters the tensor only through the product $\overrightarrow{v}\overrightarrow{v}^{T}$, which is blind to the sign. Suppose one nevertheless chooses an arrow along an axis and carries it continuously while the tensor is steered around a closed loop of control parameters (see Fig.~\ref{fig:holonomy}). If the loop encloses a degeneracy point without crossing it, the arrow can return pointing the opposite way. This sign reversal was first noticed for electronic wavefunctions near intersecting molecular energy surfaces \cite{von_Neumann_J_1929,Longuet_Higgins_H_C_1958,Herzberg_G_1963}, and later understood as the real-symmetric counterpart of the Berry phase \cite{Berry_M_V_1984}. This reversal defines a binary return parity $w_1 \in \{0,1\}$ for the principal axis being followed. The parity cannot change under a smooth deformation of the loop unless an eigenvalue degeneracy is crossed. Mathematically, $w_1$ is the first Stiefel–Whitney invariant of that headless principal axis \cite{von_Neumann_J_1929,Herzberg_G_1963,Ahn_J_2019}. Crucially, the tensor itself remains smooth and nondegenerate everywhere on the loop and the obstruction concerns only the continuous choice of an arrow along the headless principal axis. Any real symmetric rank-2 observable reported through principal values and principal axes carries the same structure of strata, seams, and sign-ambiguous axes. That ambiguity has a classical consequence for tensor fields over physical space: the index of a degeneracy is half-integral rather than whole~\cite{Berry_Hannay_1977,Delmarcelle_T_1994}.

In electronic band theory, this classification acts on Bloch wavefunctions over crystal-momentum space~\cite{Ahn_J_2019}. The same invariant is at work here, on the eigenframe of a single local tensor measured at one probe site and transported through a controlled parameter space. It emerges only when that tensor is followed around a closed loop, and requires neither quantum phase coherence nor a spatially extended state. The hyperfine electric field gradient (EFG) is a local observable of exactly this kind~\cite{Schell_J_2017,Jerschow_A_2005,Liu_Y_S_1984}: traceless by construction, measured at a single probe site, and steerable through its five-dimensional tensor space by symmetry-adapted strain modes, with the enclosing loop imposed externally.

Using first-principles strain trajectories, we establish rutile TiO$_2$ as a prototype exhibiting local chart anomalies and a global sign-reversal loop, illustrate in SnO$_2$ that degeneracies can appear as points or lines depending on the chosen control plane, and demonstrate in MgO complete local control of the traceless tensor space.


\section{Principal-axis charts for electric field gradients}\label{sec2}

\subsection{The ordered chart}
The electric-field-gradient (EFG) tensor $\overleftrightarrow{V}\in X=\operatorname{Sym}_0(3,\mathbb{R})$ at the probe site $\overrightarrow{r_0}$ is the symmetric traceless part of the Hessian of the electrostatic potential $\Phi$~\cite{Schatz_G_1996,Valenzuela_R_J_2025}:
\begin{equation}
    V_{ij}
    =
    \left[
        \partial_i\partial_j\Phi
        -
        \frac{1}{3}\delta_{ij}\nabla^2\Phi
    \right]_{\overrightarrow{r}=\overrightarrow{r_0}}.
    \label{eq:EFG_definition}
\end{equation}
Symmetry follows from the equality of mixed partial derivatives, whereas tracelessness follows by construction. In the nuclear multipole expansion, the trace enters the monopole term as the isomer shift, leaving the traceless part as the quadrupole coupling. Consequently, $\overleftrightarrow{V}$ has five independent components. Throughout, the principal values of $\overleftrightarrow{V}$ are the eigenvalues $\lambda_{i}$ of its matrix representation. We use ``principal values'' in physical discussion and ``eigenvalues'' in
formal statements. 

Since $\overleftrightarrow{V}$ is real and symmetric, it admits an orthonormal eigenbasis, the principal-axis system (PAS): there exists $\overleftrightarrow{R}=(\overrightarrow{v_1},\overrightarrow{v_2},\overrightarrow{v_3})\in \mathrm{SO}(3,\mathbb{R})$, with columns the EFG eigenvectors, such that $\operatorname{diag}(V_{11},V_{22},V_{33})=\overleftrightarrow{R}^{T}\overleftrightarrow{V}\overleftrightarrow{R}$. The principal values are conventionally ordered by their absolute magnitudes,
\begin{equation}
    |V_{33}| \geq |V_{22}| \geq |V_{11}|,
    \label{eq:V_convention}
\end{equation}
and the tensor is reported through the largest component $|V_{33}|$ and the asymmetry parameter~\cite{Schatz_G_1996,Valenzuela_R_J_2025,Kaufmann_E_N_1979,Harris_R_K_2008}
\begin{equation}
    \eta
    =
    \frac{V_{11} - V_{22}}{V_{33}},
    \label{eq:eta}
\end{equation}
which satisfies
\begin{equation}
    0 \leq \eta \leq 1.
\end{equation}

The pair $(|V_{33}|,\eta)$ retains the sorted spectrum but discards the orientation encoded in $\overleftrightarrow{R}$; it is the chart introduced in Section~\ref{sec1}, and its two distinguished loci are characterized next.

\subsection{The degeneracy stratum and ordering seam}
The degeneracy stratum $\Sigma_0\subset X$ consists of tensors for which two or more principal values coincide. Away from the origin, magnitude ordering allows the repeated values to be written as $V_{11}=V_{22}=\lambda$ with $\lambda\neq0$. Tracelessness then gives $(V_{11},V_{22},V_{33})=(\lambda,\lambda,-2\lambda)$, consistent with Eq.~\eqref{eq:V_convention}. Thus, for $\overleftrightarrow{V}\neq\overleftrightarrow{0}$, $\overleftrightarrow{V}\in\Sigma_0 \Longleftrightarrow \eta=0$ (Eq.~\ref{eq:eta}). On $\Sigma_0\setminus\{\overleftrightarrow{0}\}$, the repeated eigenvalue has a two-dimensional eigenspace: any orthonormal pair within it is an equally valid choice of axes, and the chart loses the corresponding frame directions. As the equality of two eigenvalues is a property of $\overleftrightarrow{V}$ itself, $\Sigma_0$ is independent of the chosen basis and reporting convention.

The determinant-zero set $\Delta_0\subset X$ consists of tensors for which one principal value vanishes, $\det(\overleftrightarrow{V})=V_{11}V_{22}V_{33}=0$. Under magnitude ordering, this means $V_{11}=0$, whereas $V_{22}=-V_{33}$ by tracelessness. Then, for $\overleftrightarrow{V}\neq\overleftrightarrow{0}$, $\overleftrightarrow{V}\in\Delta_0 \Longleftrightarrow \eta=1$ (Eq.~\ref{eq:eta}). Thus, $\Delta_0$ appears as an ordering seam of the reported chart: crossing it forces a branch relabelling even though the underlying tensor is generically smooth and nondegenerate.

\subsection{Relabellings and the eigenline}\label{sec2.3}
At $\Delta_0$, $V_{11}$ vanishes and $|V_{22}|=|V_{33}|$, so crossing exchanges
the labels $V_{22}$ and $V_{33}$, producing the cusp-like seam at $\eta = 1$ even
though the tensor and its principal values remain smooth and nondegenerate. Around a closed path encircling $\Sigma_0$, the tensor returns to itself while the continuously transported frame may return only up to signs, for example

\begin{equation}
    (\overrightarrow{v}_1, \overrightarrow{v}_2, \overrightarrow{v}_3)
\longrightarrow
(-\overrightarrow{v}_1, \overrightarrow{v}_2, -\overrightarrow{v}_3).
\label{eq:signflip}
\end{equation}
No measurement of the tensor distinguishes these frames. The spectral decomposition
\begin{equation}
    \overleftrightarrow{V}
    =\sum_{i=1}^{3}\lambda_i\,\overrightarrow{v}_i\overrightarrow{v}_i^{\,T},
    \qquad
    (-\overrightarrow{v}_i)(-\overrightarrow{v}_i)^{T}=\overrightarrow{v}_i\overrightarrow{v}_i^{\,T},
    \label{eq:projector}
\end{equation}
shows that the tensor is built from the sign-free projectors of its eigenlines. Each PAS direction is an unoriented eigenline, thus the EFG is naturally described by headless principal axes. Choosing a sign in Eq.~\eqref{eq:signflip} amounts to placing an arrow on a headless axis. Such a choice is arbitrary pointwise, yet a continuous lift along a path is determined by the sign chosen at its starting point, and around a closed loop, it may return reversed. As formalized in the next section, this defines a topologically nontrivial holonomy.

\newpage

\section{Spectral-geometric framework}\label{sec3}

We now work directly in the invariant tensor space $X=\mathrm{Sym}_0(3,\mathbb{R})$, equipped with the Frobenius inner product $\langle A,B\rangle_F=\operatorname{tr}(A^{\top}B)$, and study the EFG as a function of one or two control parameters chosen from symmetry-adapted strain modes of the crystal. Any symmetric strain tensor decomposes into an isotropic trace component and a traceless deviatoric component; the latter, expanded in a symmetry-adapted basis, provides the local control directions used to navigate $X$. In this intrinsic formulation, chart anomalies can be described without committing to a particular ordering convention, connecting the Herzberg--Longuet-Higgins holonomy structure of real-symmetric eigenproblems~\cite{Longuet_Higgins_H_C_1958,Herzberg_G_1963,Berry_M_V_1984} with the shape-space description familiar from traceless order parameters and deviatoric strain~\cite{de_Gennes_P_G_1993,Beris_A_N_1994}.

For a fully geometric interpretation of the EFG tensor, the ordered PAS parameters introduced in section~\ref{sec2} are insufficient, since they depend on a particular eigenvalue ordering and principal-axis convention. Instead, the relevant basis-independent quantities are the polynomial invariants of $\overleftrightarrow{V}$. 

The quadratic invariant
\begin{equation}
I_2 =
\operatorname{Tr}\left(\overleftrightarrow{V}^{\,2}\right)
=
\left\| \overleftrightarrow{V} \right\|_F^2,
\end{equation}
gives the overall basis-independent magnitude of the EFG tensor, while the cubic invariant~\cite{de_Gennes_P_G_1993,Beris_A_N_1994}
\begin{equation}
I_3 =
\operatorname{Tr}\left(\overleftrightarrow{V}^{\,3}\right)
=
3\det\left(\overleftrightarrow{V}\right),
\end{equation}
encodes the signed shape of its eigenvalue spectrum. In particular, the determinant-zero set $\Delta_0$ is characterized by $I_3=0$ and in the magnitude-ordered PAS chart, this set appears as the ordering seam $\eta=1$.

For $\overleftrightarrow{V}\neq\overleftrightarrow{0}$, it is useful to introduce the dimensionless normalized shape parameter
\begin{equation}
p =
\sqrt{6}\,
\frac{I_3}{I_2^{3/2}}.
\label{eq:p}
\end{equation}
The prefactor $\sqrt{6}$ normalizes the axial limits to $p=\pm1$. For an axial spectrum $(\lambda,\lambda,-2\lambda)$, one has $I_2=6\lambda^2$ and $I_3=-6\lambda^3$, so $p=-\operatorname{sgn}(\lambda)$. Thus, $p$ describes the signed spectral shape independently of tensor magnitude and basis: for nonzero tensors, $p=\pm1$ on the degeneracy stratum $\Sigma_0$ $(\eta=0)$, while $p=0$ on the determinant-zero set $\Delta_0$, which forms the $\eta=1$ ordering seam of the magnitude-ordered chart. Other conventions place the seam elsewhere, but $\Sigma_0$ and the return parity of a continuously tracked eigenline are properties of the tensor and not of the chart.

The two loci govern different regularity domains: the eigenframe is well defined on $X\setminus\Sigma_0$, whereas the ordered chart is smooth only on $X\setminus(\Sigma_0\cup\Delta_0)$. Local chart anomalies arise when a trajectory approaches or crosses either locus (the chart-wall form, I); global holonomy arises only when a closed loop links the relevant component of $\Sigma_0$ (the holonomy form, II).

\newpage

\paragraph{I) The chart-wall form: local singular geometry.}
Let $y\mapsto \overleftrightarrow{V}(y)\in X$ be a smooth tensor trajectory. Since $I_2$ and $I_3$ are polynomial invariants of $\overleftrightarrow{V}$, they vary smoothly along the trajectory. Nevertheless, the ordered PAS coordinates $(|V_{33}|,\eta)$ can exhibit sharp, convention-induced features when the trajectory approaches or crosses either of the two distinguished loci, $\Sigma_0$ or $\Delta_0$. We call these loci chart walls because they map to the boundaries of the chart's range, where $\eta$ attains a non-smooth extremum: a minimum of $0$ at $\Sigma_0$ and a maximum of $1$ at $\Delta_0$.

\begin{figure}[H]
    \centering
    \includegraphics[width=0.8\linewidth]{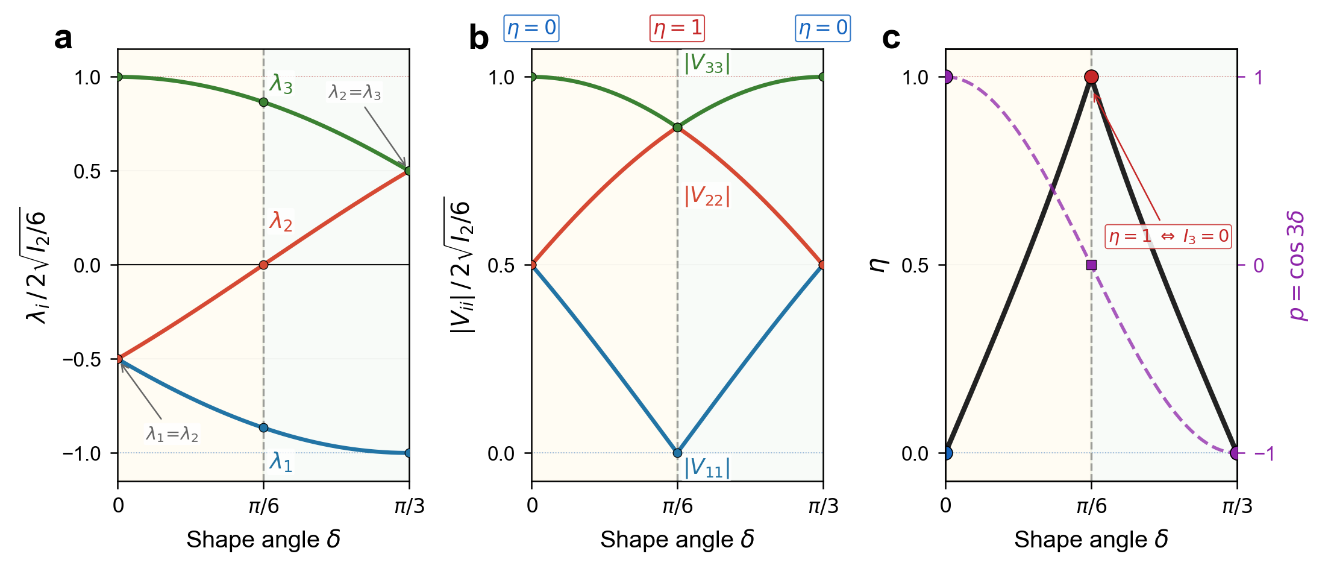}
    \caption{
    \textbf{Local chart-wall structure of the ordered PAS representation at fixed tensor magnitude $I_2$.}
    (a) The normalized principal values vary smoothly as a function of the invariant shape angle $\delta$ (see Eq.~\ref{eq:p_delta}). Here, the branches are the signed eigenvalues, sorted as $\lambda_1 \le \lambda_2 \le \lambda_3$, a different convention from the magnitude ordering of Eq.~\ref{eq:V_convention} used in (b). The endpoints $\delta=0$ and $\delta=\pi/3$ lie on the degeneracy stratum $\Sigma_0$, where two principal values coincide, and the ordered PAS parameter satisfies $\eta=0$. At $\delta=\pi/6$, one principal value vanishes ($\lambda_2=0$), corresponding to the ordering seam $\Delta_0$. (b) After imposing the ordered PAS convention, the smooth spectrum is folded into piecewise branches for $|V_{11}|,|V_{22}|,|V_{33}|$. The magnitude ordering $|V_{33}|\ge|V_{22}|\ge|V_{11}|$ forces branch reassignments at the chart walls. In particular, crossing $\Delta_0$ gives $|V_{11}|=0$ and $|V_{22}|=|V_{33}|$, producing the $\eta=1$ seam; note that the vanishing eigenvalue is $\lambda_2$ in the signed ordering of (a) but is relabelled $V_{11}$ here, and the coincident pair corresponds to $\lambda_1$ and $\lambda_3$. Approaching $\Sigma_0$ drives the spectrum to the axial limit, where two principal values coincide ($|V_{11}|=|V_{22}|$) and $\eta=0$. At this boundary the corresponding two-dimensional eigenspace is degenerate, so the associated PAS directions are no longer uniquely defined. (c) Comparison between the ordered PAS asymmetry parameter $\eta$ and the smooth invariant shape coordinate $p=\cos(3\delta)$. While $p$ varies smoothly from $+1$ to $-1$, the parameter $\eta$ is folded between the axial boundary $\eta=0$, governed by $\Sigma_0$, and the ordering seam $\eta=1$, governed by $\Delta_0$. This shows that $(|V_{33}|,\eta)$ is a convenient ordered chart, but not a globally smooth geometric representation of the EFG tensor space.}
    \label{fig:wall_form}
\end{figure}

To visualize this local chart-wall structure, we restrict to a fixed tensor magnitude $I_2$. Since the tensor is traceless, fixing $I_2$ leaves only one independent degree of freedom in the spectral shape. We parameterize this degree of freedom by an angular coordinate, defined through eq.~\ref{eq:p}:
\begin{equation}
p = \cos(3\delta),
\label{eq:p_delta}
\end{equation}
The coordinate $\delta$ provides a smooth invariant description of the eigenvalue spectrum, while the reported PAS coordinates reveal how $\eta$ approaches the chart boundaries $\eta=0$ and $\eta=1$ near $\Sigma_0$ and $\Delta_0$ respectively, see Fig.~\ref{fig:wall_form}.

\paragraph{II) The holonomy form: global topology.}
Let $C$ be a closed control loop with $\overleftrightarrow{V}(C)\subset X\setminus\Sigma_0$. Such a \emph{spectrally gapped} loop contains no principal-value degeneracy, so each axis can be followed continuously. We denote the return parity of one tracked axis by $w_1(C)$; the complete biaxial frame has richer topology.

If a spectrally gapped loop $C$ links the relevant component of $\Sigma_0$, a chosen arrow along the principal axis can be transported continuously, yet return reversed after one circuit. This gives the nontrivial parity $w_1(C)=1$, even though the tensor remains nondegenerate everywhere on $C$: the obstruction is global and arises because the loop links a degeneracy where the axis itself becomes undefined. By contrast, crossing $\Delta_0$ leaves the principal values and eigenlines smooth, but forces the magnitude-ordering convention to exchange their labels, producing a local cusp-like feature at $\eta=1$ without changing the eigenline topology. Thus, $\Sigma_0$, rather than $\Delta_0$, controls the global return parity shown in Fig.~\ref{fig:holonomy}.

\newpage
\section{Isolated eigenframe holonomy in rutile TiO\texorpdfstring{$_2$}{2}}

For the Ti site in rutile TiO$_2$, the symmetry-adapted controls $Q_{B_{1g}}$ and $Q_{B_{2g}}$ define a two-dimensional control slice in $X=\mathrm{Sym}_0(3,\mathbb{R})$ passing through the unstrained DFT EFG tensor, which has $xy \oplus z$ block form:

\begin{equation}
\overleftrightarrow{V}_0 =
\begin{pmatrix}
V_{xx,0} & V_{xy,0} &0\\
V_{xy,0} & V_{yy,0} &0\\
0&0&V_{zz,0}
\end{pmatrix},
\label{eq:Rutile_EFG}
\end{equation}
with $V_{xx,0}+V_{yy,0}+V_{zz,0}=0$. The applied strain tensor is written as 
\begin{equation}
\overleftrightarrow{\varepsilon}(Q_{B_{1g}},Q_{B_{2g}})
=
Q_{B_{1g}}\,\widehat{\varepsilon}_{B_{1g}}
+
Q_{B_{2g}}\,\widehat{\varepsilon}_{B_{2g}},
\end{equation}
where
\begin{equation}
\widehat{\varepsilon}_{B_{1g}} =
\begin{pmatrix}
1 & 0 & 0\\
0 & -1 & 0\\
0 & 0 & 0
\end{pmatrix},
\qquad
\widehat{\varepsilon}_{B_{2g}} =
\begin{pmatrix}
0 & 1 & 0\\
1 & 0 & 0\\
0 & 0 & 0
\end{pmatrix}.
\end{equation}

For each point $(Q_{B_{1g}},Q_{B_{2g}})$ in this strain-control plane, the DFT calculation defines an EFG response

\begin{equation}
\overleftrightarrow{V}
=
\overleftrightarrow{V}(Q_{B_{1g}},Q_{B_{2g}})
\in X.
\end{equation}

To leading order about the unstrained structure,
\begin{equation}
\begin{split}
\overleftrightarrow{V}(Q_{B_{1g}},Q_{B_{2g}})
={}&
\overleftrightarrow{V}_0
+
Q_{B_{1g}}
\left.
\frac{\partial\overleftrightarrow{V}}
{\partial Q_{B_{1g}}}
\right|_{(Q_{B_{1g}},Q_{B_{2g}})=(0,0)} \\
&+
Q_{B_{2g}}
\left.
\frac{\partial\overleftrightarrow{V}}
{\partial Q_{B_{2g}}}
\right|_{(Q_{B_{1g}},Q_{B_{2g}})=(0,0)}
+
\mathcal{O}\bigl(\|(Q_{B_{1g}},Q_{B_{2g}})\|^{2}\bigr).
\end{split}
\end{equation}
Here, $Q_{B_{1g}}$ and $Q_{B_{2g}}$ are scalar strain-mode amplitudes, and the derivatives give the first-order EFG change per unit amplitude of the corresponding mode.

\begin{figure}[H]
    \centering
    \includegraphics[
        width=\textwidth,
        height=0.58\textheight,
        keepaspectratio]{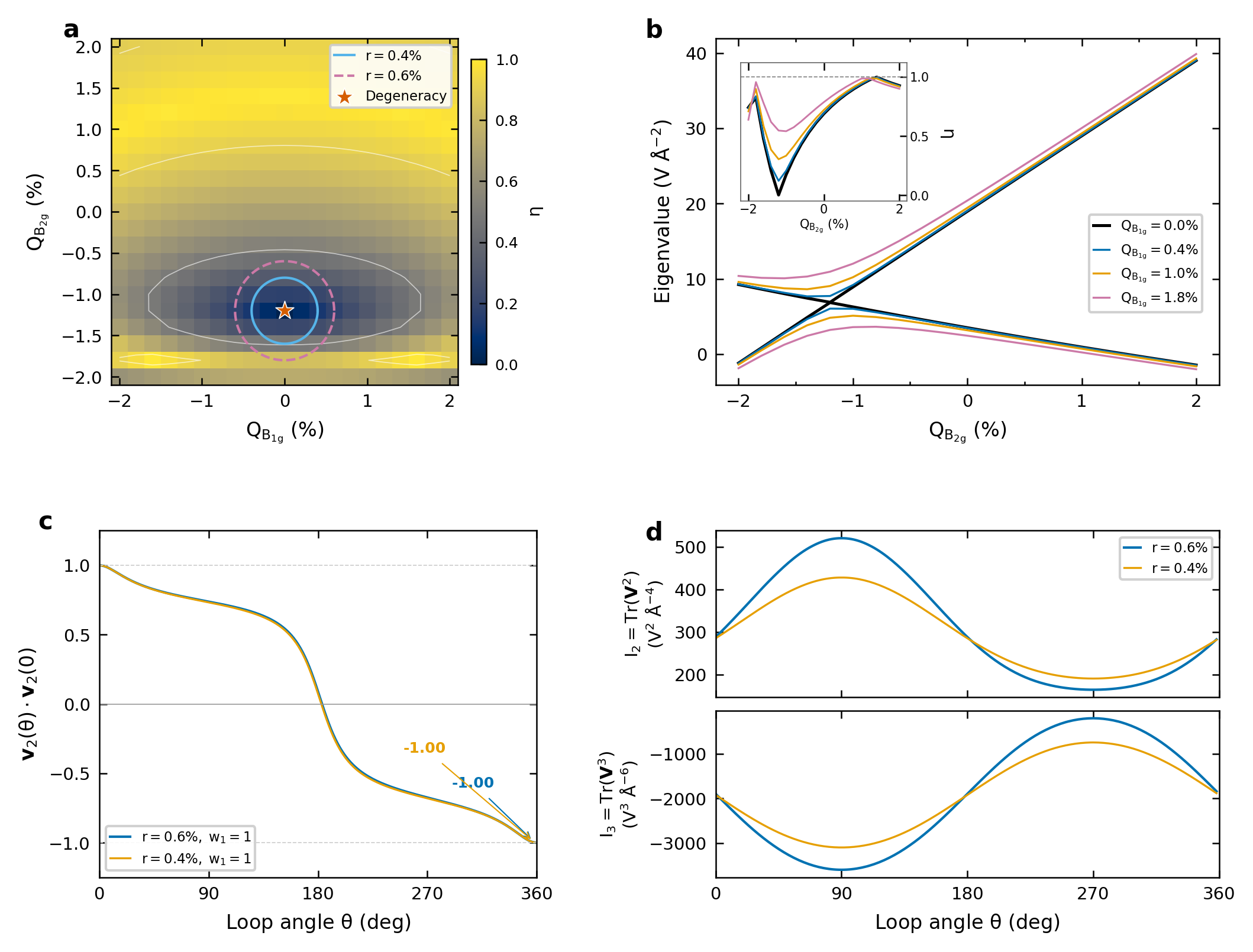}
    \caption{
    \textbf{Return parity of the EFG eigenframe in the $Q_{B_{1g}}\,$-$\,Q_{B_{2g}}$ strain-control plane of rutile TiO$_2$.}
    (a) The colour map shows the asymmetry parameter $\eta$. A point degeneracy of two principal values (star) is reached under the symmetry-adapted strains. Two closed, spectrally gapped loops $C$ of different radii are shown in blue and dashed orange. Locally, $\eta$ indicates proximity to the degeneracy, while the loop invariant $w_1(C)$ detects whether the closed path encircles it; smooth deformations of a gapped, encircling loop do not change $w_1(C)$. 
    (b) EFG principal values along cuts at several fixed values of $Q_{B_{1g}}$. Near the degeneracy, two principal-value branches approach and exchange character, the local chart-wall signature of the degeneracy stratum $\Sigma_0$; the inset shows the corresponding dip of $\eta$ toward the axial limit $\eta=0$. 
    The avoided crossings seen in one-dimensional cuts arise from an isolated conical intersection, or diabolical point, in the full control plane. The local structure resembles a tilted Dirac cone, except that the surfaces represent EFG principal values rather than electronic bands.
    (c) Normalized overlap $\vec{v}_2(\theta)\cdot\vec{v}_2(0)$, where $\vec{v}_2$ is a chosen arrow along the selected principal axis. After one full circuit, $\theta=2\pi$, the overlap approaches $-1$ rather than $+1$: the arrow returns reversed, $w_1(C)=1$. The same behaviour for both loop radii confirms the topological robustness of the parity.
    (d) Basis-independent invariants $I_2$ and $I_3$ along the same loops. Both vary smoothly, confirming that the sign reversal is not caused by any singularity of the underlying tensor trajectory, but by the global topology of eigenframe transport around the degeneracy stratum.}
    \label{fig:TiO2}
\end{figure}

In this slice, the map of $\eta$ over the control plane (Figs.~\ref{fig:holonomy} and \ref{fig:TiO2}a) reveals an isolated point where two principal values coincide. Fits to the Cartesian EFG components show that $V_{xx}=V_{yy}$ and $V_{xy}=0$ at a single strain point. The nonzero Jacobian there confirms that the $Q_{B_{1g}}\,$-$\,Q_{B_{2g}}$ strain plane crosses the degeneracy stratum $\Sigma_0$ transversely (Methods E4). This single point supplies both forms of Section~3. Locally, it is a conical intersection: one-dimensional cuts show principal-value branches that exchange character, with avoided crossings on cuts that miss the point and the axial limit $\eta=0$ on the cut through it (Fig.~\ref{fig:TiO2}b). Globally, a cut \emph{through} $\Sigma_0$ carries no invariant, because the frame is singular at the degeneracy itself. The return parity becomes well defined once the cut is opened into a closed, spectrally gapped loop $C$ around the point. The two enclosing loops shown in Fig.~\ref{fig:TiO2}a are followed explicitly in Fig.~\ref{fig:TiO2}c, where the normalized eigenvector overlap approaches $-1$ after one complete circuit. The invariants $I_2$ and $I_3$ remain smooth along the same loops (Fig.~\ref{fig:TiO2}d), showing that the orientation reversal is global while the tensor trajectory remains spectrally gapped. Thus, rutile TiO$_2$ realizes both the chart-wall form and the holonomy form from first principles.

\newpage
\section{Slice-dependent degeneracy structure in SnO\texorpdfstring{$_2$}{2}}
\begin{figure}[H]
    \centering
    \includegraphics[width=\textwidth]{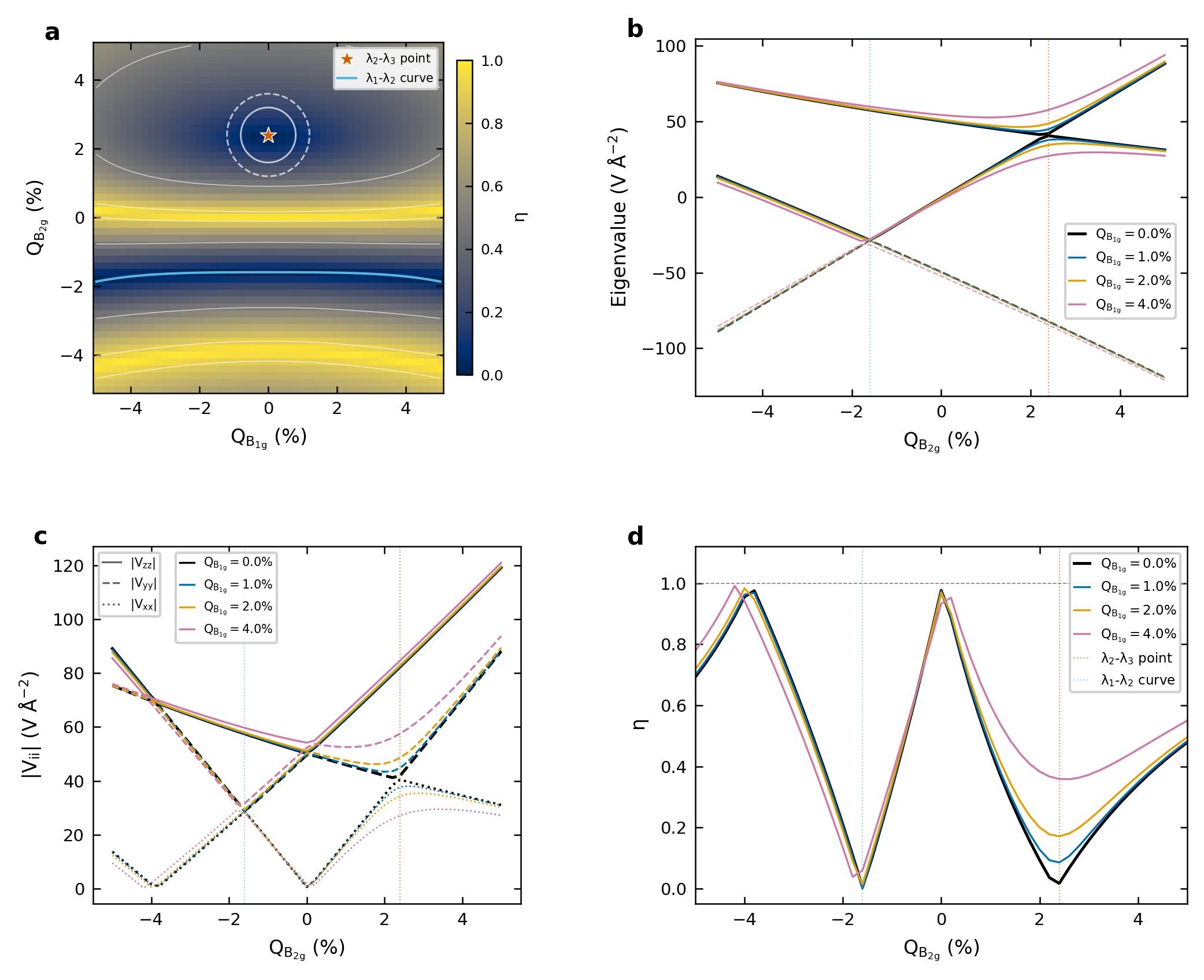}
    \caption{
    \textbf{Chart-wall structure of the EFG tensor in the $Q_{B_{1g}}\,$-$\,Q_{B_{2g}}$ strain-control plane of rutile SnO$_2$.}
    (a) Colour map of the asymmetry parameter $\eta$. The star marks an isolated $\lambda_2=\lambda_3$ degeneracy point; the blue curve marks an extended $\lambda_1=\lambda_2$ degeneracy branch. The isolated point behaves analogously to the TiO$_2$ case, whereas the extended branch shows that, in this two-dimensional control slice, the $\lambda_1=\lambda_2$ degeneracy is not fully lifted by the available strain controls. 
    (b) EFG principal values along cuts at several fixed values of $Q_{B_{1g}}$. Near the isolated point, the principal-value branches reproduce the conical-intersection structure found in TiO$_2$. In contrast, the $\lambda_1=\lambda_2$ degeneracy forms a continuous curve because the restricted control plane preserves the $xy\oplus z$ block structure. An additional $E_g$-type $xz$ or $yz$ control would introduce the off-block coupling needed to lift the crossing locally.
    (c) Ordered PAS magnitudes $|V_{11}|,|V_{22}|,|V_{33}|$ along the same cuts: the magnitude-ordering convention forces label reassignments, producing sharp, non-smooth branches even though the underlying tensor varies smoothly. 
    (d) Asymmetry parameter $\eta$ along the same cuts. Values $\eta\to0$ signal approach to the degeneracy stratum $\Sigma_0$; in contrast to the isolated point, the extended branch produces a region where $\eta$ remains pinned at zero, because the trajectory can follow the degeneracy set over a finite interval rather than crossing it at a single point. Values $\eta\to1$ signal crossing of the ordering seam $\Delta_0$, where one principal value vanishes. Overall, SnO$_2$ illustrates the distinction between isolated and extended intersections with $\Sigma_0$: isolated intersections appear as points that can carry loop parity, while extended intersections appear as lines of pinned $\eta=0$ that no spectrally gapped loop within the same two-dimensional plane can locally encircle, so the return parity is not resolved in this slice.}
    \label{fig:SnO2}
\end{figure}

Compared with TiO$_2$, the SnO$_2$ control plane displays a richer degeneracy structure. In the $Q_{B_{1g}}\,$-$\,Q_{B_{2g}}$ slice, two distinct intersections with the discriminant stratum $\Sigma_0$ are observed. The first is an isolated point degeneracy, where $\lambda_2 = \lambda_3$, analogous to the degeneracy found in TiO$_2$. The second is an extended degeneracy branch, where $\lambda_1 = \lambda_2$. Both features correspond to axially symmetric EFG tensors and thus, to the ordered PAS limit $\eta=0$, but they have different geometrical roles within the chosen two-dimensional control slice, as shown in Fig.~\ref{fig:SnO2}(a).

The extended branch is protected within the restricted $Q_{B_{1g}}\,$-$\,Q_{B_{2g}}$ control plane. These two controls preserve the $xy\oplus z$ block structure of the EFG tensor (Eq.~\ref{eq:Rutile_EFG}), so that $V_{xz}=V_{yz}=0$ and the $z$-like principal direction remains decoupled from the in-plane subspace. This symmetry-enforced decoupling allows the $z$-like and in-plane principal values to cross without level repulsion. Hence, within the two-dimensional control plane, the degeneracy condition can define a continuous line. An $E_g$-type $xz$ or $yz$ control couples the two sectors, lifts the degeneracy locally, and permits a spectrally gapped loop in a plane transverse to the degeneracy line.

\begin{figure}[H]
    \centering
    \includegraphics[width=0.8\linewidth]{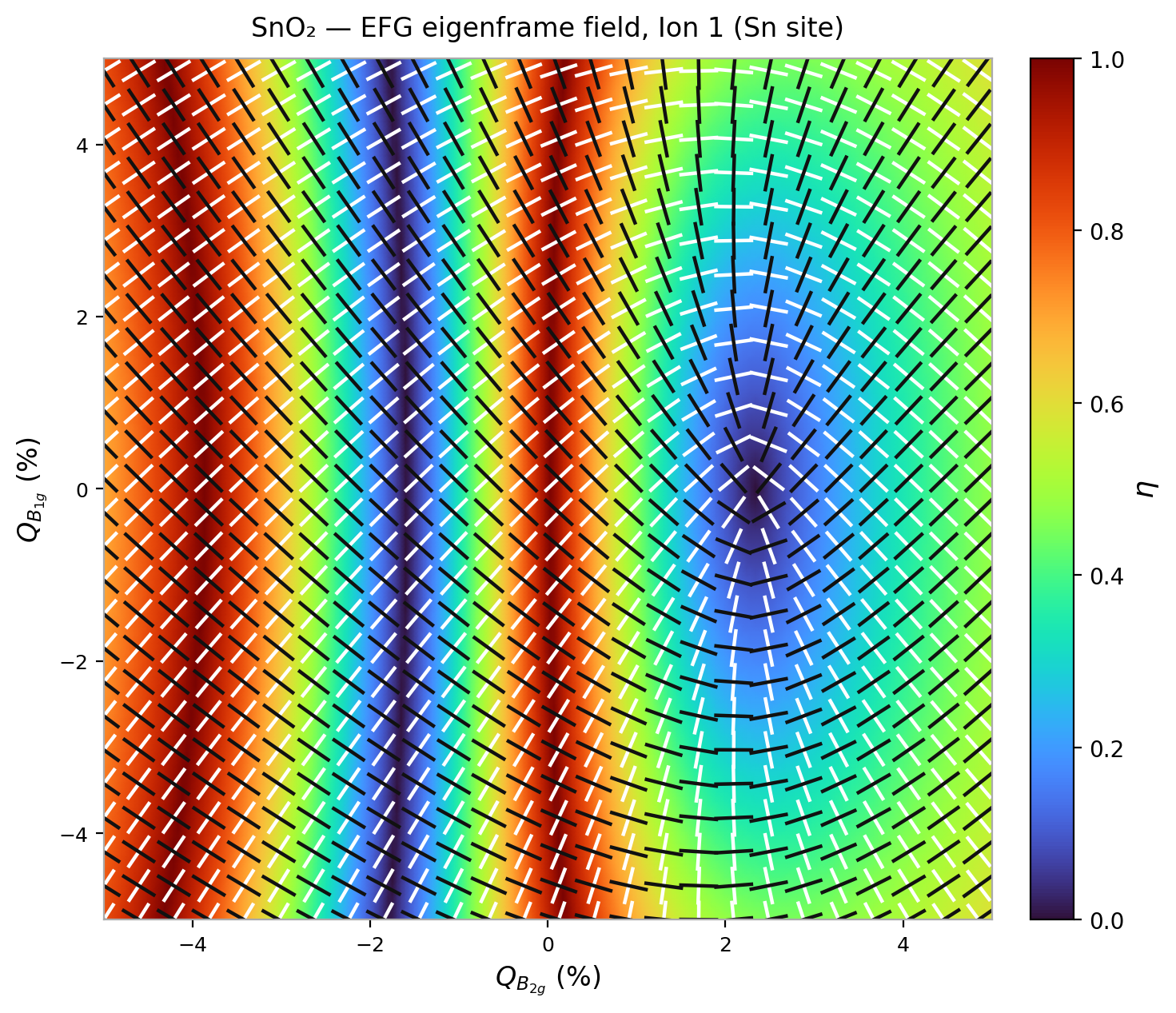}
    \caption{
    \textbf{The SnO$_2$ EFG eigenframe field in the $Q_{B_{1g}}\,$-$\,Q_{B_{2g}}$ control plane.}
    The colour map shows the asymmetry parameter $\eta$; the black and white headless line segments show the two in-plane principal axes of the EFG tensor. Both an isolated point degeneracy and an extended degeneracy branch are visible. The isolated point acts as a winding centre for the headless axes: the frame winds by half a turn per circuit, the field pattern of nontrivial return parity, $w_1=1$. In contrast, the extended branch appears as a line of singular frames along which $\eta=0$ remains pinned. Any loop attempting to encircle the branch within this two-dimensional slice must cross the degeneracy stratum $\Sigma_0$, where the frame is undefined, so no spectrally gapped loop exists around it and the return parity is not resolved in this slice. Leaving this plane through an $E_g$-type control would permit a gapped loop transverse to the branch.
    The $\eta=1$ regions are ordering seams $\Delta_0$ of the reporting convention and do not represent degeneracies of the principal values.}
    \label{fig:eigenframe_SnO2}
\end{figure}

The isolated point admits a gapped encircling loop with $w_1(C)=1$. The extended branch cannot be encircled in the same slice without crossing $\Sigma_0$, where the frame is singular, so its loop parity is unresolved there (Fig.~\ref{fig:eigenframe_SnO2}).

Thus, SnO$_2$ shows how a control slice can intersect $\Sigma_0$ as an encirclable point or a degeneracy line (Figs.~\ref{fig:SnO2} and \ref{fig:eigenframe_SnO2}).

\newpage
\section{Complete local tensor control in cubic MgO}

MgO provides a clean, high-symmetry example in which strain acts as a local control field for the EFG tensor. In the unstrained cubic reference structure, the Mg site has cubic symmetry, so the EFG vanishes, $\overleftrightarrow{V}_0=\overleftrightarrow{0}$. Consequently, the basis-independent invariants $I_2$ and $I_3$ also vanish, and the PAS is not uniquely defined at the cubic point. 

As summarized in the symmetry-adapted basis construction in Methods D, the symmetric strain tensor decomposes into an isotropic $A_{1g}$ trace component and a five-dimensional traceless deviatoric sector. The $A_{1g}$ component changes the volume but preserves the cubic site symmetry. Hence, it does not define an independent traceless EFG direction. By contrast, the $E_g$ and $T_{2g}$ deviatoric strain modes transform in the same symmetry-adapted rank-2 tensor basis as the EFG tensor. Thus, excluding the isotropic trace mode, the strain basis and the EFG basis are naturally matched by symmetry.

Near the cubic reference point, where $\overleftrightarrow{V_0}=\overleftrightarrow{0}$, the strain-induced EFG is described to leading order by the linear response:

\begin{equation}
V_{ij}=\sum_{\alpha}
\frac{\partial V_{ij}}{\partial \epsilon_\alpha}
\epsilon_\alpha
+
\mathcal{O}(\epsilon^2),
\end{equation}
where $i,j\in{x,y,z}$, and $\epsilon_\alpha$ denotes the amplitude of a symmetry-adapted strain mode. The term $\mathcal{O}(\epsilon^2)$ collects contributions of quadratic and higher order in the mode amplitudes, negligible in the small-strain regime. 

Here, the derivative $\partial V_{ij}/\partial \epsilon_\alpha$ measures the response of the Cartesian EFG component $V_{ij}$ to the mode amplitude $\epsilon_\alpha$, rather than to an individual Cartesian strain component. After excluding the isotropic $A_{1g}$ trace mode and vectorizing the five independent traceless EFG components, this response defines a $5\times5$ susceptibility matrix. The calculated matrix is full rank to numerical accuracy, indicating that the symmetry-adapted deviatoric strain modes form a complete local control basis for $X=\mathrm{Sym}_0(3,\mathbb{R})$.

The MgO results provide a constructive realization of the framework: suitable strain trajectories can target singular PAS features while locally controlling the EFG magnitude $I_2$ and spectral shape $p$ (or $\eta$).

\newpage
\begin{figure}[H]
    \centering
    \includegraphics[width=\textwidth]{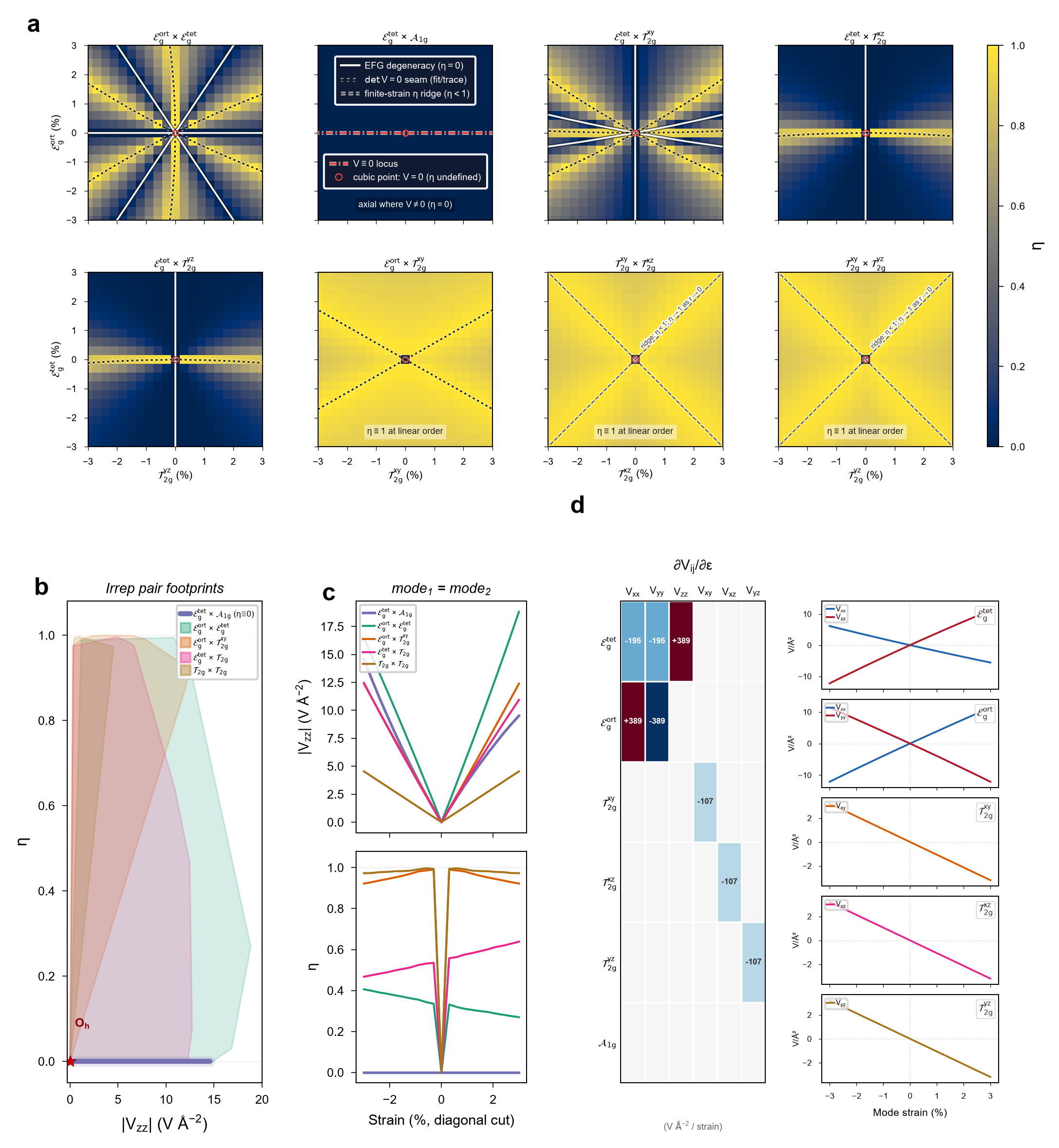}
    \caption{
    \textbf{Local strain control of the EFG tensor in cubic MgO.}
    (a) Maps of the ordered-PAS asymmetry parameter $\eta$ in representative two-dimensional slices of the strain-control space, each spanned by a different pair of symmetry-adapted strain modes. Solid white curves mark eigenvalue-degeneracy loci $\Sigma_0$, where two principal values coincide and $\eta=0$. Dark dotted curves mark the determinant-zero set $\Delta_0$, where one principal value vanishes and $\eta=1$. Red dash-dotted loci and red open markers indicate $\overleftrightarrow{V}=\overleftrightarrow{0}$, where the PAS and $\eta$ are undefined; the displayed value $\eta=0$ is a plotting convention. Grey dashed curves mark finite-strain ridges with $\eta<1$.
    (b) Footprints of the same mode pairs in the $(|V_{33}|,\eta)$ chart. The star marks the unstrained cubic structure, $\overleftrightarrow{V}=\overleftrightarrow{0}$. Different strain planes access distinct regions of the magnitude-ordered principal-axis-system chart.
    (c) Equal-amplitude cuts, $\mathrm{mode}_1=\mathrm{mode}_2$, showing $|V_{33}|$ (upper) and $\eta$ (lower) as functions of strain. The trajectories span axial, intermediate and near-maximal asymmetry. All cuts pass through $\overleftrightarrow{V}=\overleftrightarrow{0}$ at zero strain, where the plotted $\eta=0$ is conventional.
    (d) Linear response $\partial V_{ij}/\partial\epsilon_\alpha$ of the Cartesian EFG components to each symmetry-adapted mode amplitude. The two $E_g$ modes span the traceless diagonal sector, the three $T_{2g}$ modes control the off-diagonal components, and $A_{1g}$ gives no linear EFG response. Excluding $A_{1g}$, the $5\times5$ response matrix is full rank, establishing complete local linear control of $\mathrm{Sym}_0(3,\mathbb{R})$.}
    \label{fig:MgO}
\end{figure}

\newpage
\section{Measuring return parity}

Return parity is accessible to existing hyperfine methods. Multi-axis piezoelectric strain control can generate closed control loops~\cite{Hicks_C_W_2014,Barber_M_E_2019}, while testing return parity requires orientation-resolved principal-axis projectors, in addition to the spectral parameters $(|V_{33}|,\eta)$. Fixed-geometry single-crystal TDPAC can resolve the orientation discarded by powder averaging~\cite{Yap_I_C_J_2022,PRB_paper}, as can single-crystal NMR rotation patterns~\cite{Fojud_Z_2007,Kissikov_T_2017,Kawamura_M_2010}. Likewise, angle-dependent single-crystal M\"ossbauer line intensities determine the EFG axes in $^{119}$Sn absorbers~\cite{Szymanski_K_2018,Negita_H_1977}. A complementary native-site route is thermal-neutron activation of enriched $^{118}$SnO$_2$, which produces $^{119m}$Sn directly on the host Sn sublattice~\cite{Hannaford_P_1965,Hannaford_P_1969}. In a thin single-crystal specimen, orientation-resolved e$^-$--$\gamma$ TDPAC could exploit the strongly converted $^{119m}$Sn cascade~\cite{Schell_J_2017,Soares_J_C_1973} to follow the Sn-site EFG eigenframe under strain. In either implementation, an additional $E_g$-type strain could define a spectrally gapped loop transverse to the protected degeneracy branch.

Operationally, one reconstructs the principal-axis projector $P_i=\overrightarrow{v}_i\overrightarrow{v}_i^{\,T}$ at each point of an enclosing and a non-enclosing control loop. Each spectrum fixes a headless axis, and it is experimentally convenient to fit a single signed direction; the opposite sign is an equally valid representative. The parity is read from the continuous lift of Sec.~\ref{sec2.3}: the sign is fixed arbitrarily at the starting point, and at each subsequent point, the sign is chosen to give positive overlap with the preceding one. This rule introduces no further arbitrariness, since a fit returning the opposite direction is simply flipped back. It tracks the true continuation provided the axis rotates by less than $\pi/2$ between neighbouring points, which sets the required sampling density where the eigenframe turns fastest. The lift returns reversed only for the enclosing loop (Fig.~\ref{fig:holonomy}). Reversing the initial sign reverses the entire lift, leaving the parity unchanged. Therefore, per-point angular accuracy is not itself limiting: the reconstruction fails only if fit uncertainty carries a single increment past $\pi/2$, so traversing the loop in both directions and at two radii verifies that no such misassignment occurs.

\backmatter

\bmhead{Data availability}

The first-principles data supporting this study, including the calculated EFG tensors, strain-control trajectories, eigenvalue/eigenframe data, and raw DFT output files, will be deposited in a public repository before publication. Repository accession information and persistent identifiers will be added to the final version of the manuscript. Data are available from the corresponding author upon reasonable request during review.

\bmhead{Supplementary information}

This manuscript is intended to be self-contained. No supplementary information is provided at this stage.



\bmhead{Acknowledgements}

Financial support was provided by the Federal Ministry of Research, Technology, and Space (BMFTR) through Grants No. 05K16PGA, 05K22PGA, 05K25PGA, and 05K22PGB; by the ISOLDE Collaboration; by the EU Horizon Europe Programme under Grant Agreement No. 101057511 (EURO-LABS).

Computational resources were provided by the Polish PLGrid Infrastructure through ACK Cyfronet AGH under Grant No. PLG/2023/016756. The authors also acknowledge computing time granted by the Center for Computational Sciences and Simulation (CCSS) of the University of Duisburg-Essen and provided on the supercomputer amplitUDE at the Zentrum für Informations- und Mediendienste (ZIM), funded through DFG project 459398823, grant ID INST 20876/423-1 FUGG.



\section*{Declarations}

\bmhead{Competing interests}

The authors declare no competing interests.

\bmhead{Author contributions}

I.C.J.Y., S.Q.J., B.D. contributed equally to this work. Co-first author ordering is in descending order of seniority. H.C.H., D.C.L., A.K. and J.H.S. supervised this work equally. Co-senior author ordering was determined by mutual agreement.

I.C.J.Y. conceived the spectral-geometric and topological framework, developed the theoretical interpretation, performed the eigenframe-holonomy analysis, prepared the manuscript structure, and wrote the original draft. S.Q.J. performed the first-principles calculations, developed and implemented the strain-control workflows, analysed the calculated EFG tensors, and contributed to data visualization and interpretation. B.D. contributed to the conceptual development, theoretical analysis, figure preparation, and interpretation of the tensor-space and eigenframe-topology results.

T.T.D. contributed to the hyperfine-spectroscopy interpretation, validation of the physical setting, and critical revision of the manuscript. P.M.S. contributed to discussion of the hyperfine-spectroscopy context, interpretation of local EFG observables, and critical revision of the manuscript.

H.C.H., D.C.L., A.K. and J.H.S. supervised the project, contributed to interpretation of the results, provided scientific guidance, and revised the manuscript critically. All authors discussed the results and approved the final manuscript.









\begin{appendices}



\section*{Methods}

\subsection*{(A) Invariant parametrization of traceless EFG spectra}

We represent the EFG at the probe site by a real symmetric traceless tensor $\overleftrightarrow{V} \in X = \operatorname{Sym}_{0}(3,\mathbb{R})$, and characterize its spectrum using the quadratic and cubic rotation invariants:
\begin{equation}
I_{2} = \operatorname{Tr}(\overleftrightarrow{V}^{2}), \qquad
I_{3} = \operatorname{Tr}(\overleftrightarrow{V}^{3}).
\tag{M1}
\end{equation}
For a traceless $3 \times 3$ tensor, $I_{3} = 3\det(\overleftrightarrow{V})$.

The characteristic polynomial can be written in the invariant form as:
\begin{equation}
\chi_{\overleftrightarrow{V}}(\lambda)
= \det(\lambda \overleftrightarrow{Id} - \overleftrightarrow{V})
= \lambda^{3} - \frac{I_{2}}{2}\lambda - \frac{I_{3}}{3},
\tag{M2}
\end{equation}
so the principal values $\{\lambda_i\}_{i\in\{1,2,3\}}$ are the real roots of the depressed cubic.

To detect eigenvalue degeneracies, we use the polynomial discriminant of the characteristic cubic:
\begin{equation}
\operatorname{disc}\overleftrightarrow{V}
= \frac{1}{2}I_{2}^{3} - 3I_{3}^{2}
= \prod_{i<j}(\lambda_i - \lambda_j)^2,
\tag{M3}
\end{equation}
which vanishes if and only if two eigenvalues coincide.

This singles out the discriminant, or degeneracy, stratum:
\begin{equation}
\Sigma_{0} := \{\overleftrightarrow{V}\in X \mid \operatorname{disc}\overleftrightarrow{V}=0\}
\equiv \{\overleftrightarrow{V}\in X \mid \exists i\neq j,\lambda_i=\lambda_j\},
\tag{M4}
\end{equation}
and the determinant-zero set:
\begin{equation}
\Delta_{0} := \{\overleftrightarrow{V}\in X \mid \det\overleftrightarrow{V}=0\}
\equiv \{\overleftrightarrow{V}\in X \mid \exists i,\lambda_i=0\}.
\tag{M5}
\end{equation}
We note that the origin $\overleftrightarrow{V}=\overleftrightarrow{0}$ is the unique triple-degenerate point of $\Sigma_0$ and the unique intersection of $\Sigma_0$ with $\Delta_0$. Elsewhere on $\Sigma_0$, the degeneracy is twofold, whereas elsewhere on $\Delta_0$, the spectrum is nondegenerate.

The set $\Sigma_0$ is intrinsic to tensor space because it is defined by the equality of two eigenvalues, independently of the chosen basis or principal-value ordering. Likewise, the set $\Delta_0$ is defined invariantly, but its role as an ordering seam arises only after the magnitude-ordering convention is imposed. A generic twofold degeneracy of a real-symmetric tensor requires two independent conditions. This means that locally, the two diagonal entries of the participating two-level block must become equal and its off-diagonal term must vanish (see Methods C). Consequently, the intersection with $\Sigma_0$ appears as an isolated point in a generic two-parameter control plane. However, within a symmetry-restricted control space, the off-diagonal term may vanish identically, leaving only the equality of the diagonal entries to be tuned and allowing an extended degeneracy branch, as in SnO$_2$ (Fig.~\ref{fig:SnO2}). In contrast, $\Delta_0$ requires only one principal value to vanish and therefore appears generically as a curve in such a plane (Figs.~\ref{fig:TiO2},~\ref{fig:SnO2}).

For a nonzero tensor, $I_2>0$, and we define the dimensionless
invariant ratio:
\begin{equation}
p := \frac{\sqrt{6}I_{3}}{I_{2}^{3/2}} \in [-1,1],
\qquad
\delta(p) := \frac{1}{3}\cos^{-1}(p) \in \left[0,\frac{\pi}{3}\right].
\tag{M6}
\end{equation}
Non-negativity of the discriminant bounds this ratio: substituting $I_{3}^{2}=p^{2}I_{2}^{3}/6$ into Eq.~(M3) gives $\operatorname{disc}\overleftrightarrow{V}=\tfrac{1}{2}I_{2}^{3}(1-p^{2})\ge 0$, so the prefactor $\sqrt{6}$ places the bound at $|p|\le 1$, with equality attained
on $\Sigma_0\setminus\{\overleftrightarrow{0}\}$. The factor $1/3$ then maps $\cos^{-1}(p)\in[0,\pi]$ to $\delta\in[0,\pi/3]$.

The trigonometric solution for $\chi(\overleftrightarrow{V})=0$ gives via Cardano~\cite{NIST_Cubic_Equations_2025}:
\begin{equation}
\lambda_{k}
= 2\sqrt{\frac{I_{2}}{6}}\cos\left(\delta(p)+\frac{2\pi k}{3}\right),
\qquad k\in\{0,1,2\},
\tag{M7}
\end{equation}
and we use the principal branch $\delta\in[0,\pi/3]$ for a unique shape parameter. The trigonometric roots are obtained up to permutation; we then \textbf{sort} them to enforce the signed ordering $\lambda_{1}\leq\lambda_{2}\leq\lambda_{3}$:
\begin{equation}
\lambda_{1}=2\sqrt{\frac{I_{2}}{6}}\cos\left(\delta+\frac{2\pi}{3}\right),\qquad
\lambda_{2}=2\sqrt{\frac{I_{2}}{6}}\cos\left(\delta+\frac{4\pi}{3}\right),\qquad
\lambda_{3}=2\sqrt{\frac{I_{2}}{6}}\cos(\delta).
\tag{M8}
\end{equation}

\newpage

\subsection*{(B) Invariant mapping from invariants to the PAC chart}

Because the PAC convention orders principal values by absolute magnitude, $(|V_{33}|,\eta)$ is obtained by composing the smooth invariant map $\overleftrightarrow{V}\rightarrow(I_{2},I_{3})$ with a piecewise PAS labelling rule. Following equations M6 to M8, the PAC largest absolute component is:
\begin{equation}
|V_{33}| := \max_{i\in\{1,2,3\}}|\lambda_i|.
\tag{M9}
\end{equation}
The principal value associated with $|V_{33}|$ switches at $\delta=\pi/6$, where the determinant-zero set becomes the ordering seam of the magnitude-ordered chart:
\begin{equation}
\delta = \frac{\pi}{6}
\Leftrightarrow |\lambda_{3}| = |\lambda_{1}|
\Leftrightarrow \lambda_{2}=0
\Leftrightarrow \det(\overleftrightarrow{V})=0
\Leftrightarrow I_{3}=0\ (p=0).
\tag{M10}
\end{equation}
A compact way to encode the folding induced by the absolute-magnitude assignment is to define the folded angle:
\begin{equation}
\delta_{*}(p)
:= \min\left\{\delta(p),\frac{\pi}{3}-\delta(p)\right\}
\in\left[0,\frac{\pi}{6}\right]
\Leftrightarrow
\delta_{*}(p)=\frac{1}{3}\cos^{-1}(|p|),
\tag{M11}
\end{equation}
so that
\begin{equation}
|V_{33}| = 2\sqrt{\frac{I_{2}}{6}}\cos(\delta_{*}).
\tag{M12}
\end{equation}
Within each PAS-ordered region one has the equivalent piecewise form
\begin{equation}
\eta =
\begin{cases}
\sqrt{3}\tan(\delta(p)), & \delta\in\left[0,\frac{\pi}{6}\right],\\[2mm]
\sqrt{3}\tan\left(\frac{\pi}{3}-\delta(p)\right), & \delta\in\left[\frac{\pi}{6},\frac{\pi}{3}\right],
\end{cases}
\nonumber
\end{equation}
which folds to a single closed expression
\begin{equation}
\eta(\overleftrightarrow{V})
= \sqrt{3}\tan\left(\frac{1}{3}\cos^{-1}(|p|)\right)
\in [0,1].
\tag{M13}
\end{equation}
The absolute value $|p|$ arises because magnitude ordering treats $\overleftrightarrow{V}$ and $-\overleftrightarrow{V}$ equivalently. Under overall sign reversal, $I_2$ remains unchanged, whereas $I_3$ (and hence $p$) changes sign, while the reported parameters $(|V_{33}|,\eta)$ remain unchanged.

This mapping makes the two loci explicit: $\eta\rightarrow 0$ at the discriminant boundary (axial limits $\delta\rightarrow 0$ or $\delta\rightarrow \pi/3$, where an eigenvalue degeneracy is approached), while $\eta\rightarrow 1$ occurs at the folding seam $\delta=\pi/6$, i.e. at $\Delta_{0}$ where one principal value crosses zero.

We note that the invariant pair underlying this chart has appeared previously in a statistical setting: in the analysis of quadrupole-splitting distributions in disordered solids, Czjzek described the folded geometry of the $(V_{33}, \eta)$ parameter space (Ref. \cite{Czjzek_G_1981,Czjzek_G_1983}), and the conformal coordinates later introduced by Evenson et al. (Ref. \cite{Evenson_W_E_2016}) are, up to normalization, $(\operatorname{disc}\overleftrightarrow{V})^{1/2}$ and $\operatorname{det}\overleftrightarrow{V}$ — the defining polynomials of $\Sigma_{0}$ and $\Delta_{0}$. 

Concretely, their map is the cubing $W=W_{1}+\mathrm{i}W_{2}=Z^{3}$ of the complex Czjzek coordinate $Z$, which unfolds the $60^{\circ}$ ordering sector into the half-plane $W_{1}\geq 0$: the two $\eta=0$ edges of the sector are carried onto the halves of the boundary $W_{1}=0$ (i.e.\ $\Sigma_{0}$), and the $\eta=1$ midline onto $W_{2}=0$ (i.e.\ $\Delta_{0}$). That sector is the fundamental domain of the eigenvalue-relabelling action, so their unfolding removes the ambiguity responsible for the chart wall at $\Sigma_0$. Instead, the ordering seam $\Delta_{0}$ arises from the residual involution $\overleftrightarrow{V}\mapsto-\overleftrightarrow{V}$, which magnitude ordering cannot resolve because it reports $|V_{33}|$; on the spectrum it acts as the reflection $W\mapsto\overline{W}$, gluing their half-plane to itself along $W_{2}=0$.

However, in the statistical setting, the orientational average removes the eigenframe, so the transport and holonomy of Methods C have no counterpart there. The seam has also surfaced recently as a computational pitfall: in quantum-chemical EFG benchmarks, small variations of the smallest principal value near $\eta\approx1$ were observed to flip the assigned sign of the largest-magnitude principal component discontinuously~\cite{Derevianko_A_2026}; in the present terms, such flips are transverse crossings of $\Delta_{0}$, Eq.~(M10), and carry no physical discontinuity.

\subsection*{(C) Two-level reduction and eigenframe transport}

When only two principal values approach degeneracy along a control trajectory, we reduce the $3\times3$ real-symmetric traceless EFG $\overleftrightarrow{V}\in X=\operatorname{Sym}_{0}(3,\mathbb{R})$ to an effective real two-level problem by isolating the participating eigenvalue pair into a symmetric $2\times2$ subblock, with the third eigenvalue treated as a spectator. This is the real (class-AI) analogue of a Berry-phase setting, where loops enclosing a degeneracy can produce a quantized sign holonomy of the transported real eigenvector~\cite{von_Neumann_J_1929,Longuet_Higgins_H_C_1958,Herzberg_G_1963,Berry_M_V_1984}.  A complementary differential-geometric treatment, endowing the space of $2\times2$ real symmetric matrices with a connection whose eigenframes are parallel and whose repeated-eigenvalue set carries the curvature, was given recently in Ref.~\cite{Rondomanski_J_2024}; here we retain the flat Frobenius structure on $X$ and extract only the gauge-intrinsic parity.

\subsubsection*{(C1) Symmetric-gauge isolation of a participating pair}

We start with a general traceless real symmetric EFG tensor $\overleftrightarrow{V}\in X=\operatorname{Sym}_{0}(3,\mathbb{R})$. Triple degeneracy requires $\overleftrightarrow{V}=\overleftrightarrow{0}$ (a consequence of imposing the condition that all eigenvalues equal, together with the traceless property), corresponding to the cubic-symmetry point. In this case, $(|V_{33}|,\eta)$ is ill-defined. Therefore, we consider cases up to double degeneracies, where only one pair of eigenvalues approach each other along the control trajectory.

\textbf{Step 1} (Diagonalize): At each control point, we compute the spectral decomposition
\begin{equation}
\boldsymbol{\Lambda}
= \operatorname{diag}(\lambda_{1},\lambda_{2},\lambda_{3})
= \mathbf{Q}^{T}\mathbf{V}\mathbf{Q},
\qquad
\mathbf{Q} :=
\begin{bmatrix}
\widehat{v}_{1} & \widehat{v}_{2} & \widehat{v}_{3}
\end{bmatrix}
\in SO(3,\mathbb{R}),
\tag{M14}
\end{equation}
where we enforce $\det(\mathbf{Q})=+1$ by a sign flip if needed. We order the signed eigenvalues as $\lambda_{1}\leq\lambda_{2}\leq\lambda_{3}$ (or another stated convention).

\textbf{Step 2} (Choosing spectator and participating subspace): We identify the spectator index $s\in\{1,2,3\}$ as the eigenvalue $\lambda_s$ that remains best separated from the other two (largest instantaneous spectral gap) along the loop and denote its eigenvector by $\widehat{v}_{s}$. The participating two-dimensional subspace is the orthogonal complement:
\begin{equation}
\mathcal{P} := \{\vec{x}\in\mathbb{R}^{3}\mid \vec{x}\cdot\widehat{v}_{s}=0\}.
\tag{M15}
\end{equation}
The participating eigenvalues are the eigenvalues of the restriction $\left.\overleftrightarrow{V}\right|_{\mathcal{P}}$, i.e. of the $2\times2$ block defined below.

\textbf{Step 3} (Fix a symmetric gauge in the participating plane): To obtain a generally non-diagonal real $2\times2$ representation needed for the winding analysis, we fix an orthonormal basis $\{\widehat{u}_{1},\widehat{u}_{2}\}$ of $\mathcal{P}$ without using the instantaneous participating eigenvectors. Concretely, choose a fixed laboratory/crystal reference unit vector $\widehat{r}$ (e.g. $\widehat{x}$) and project it into $\mathcal{P}$:
\begin{equation}
\vec{u}_{1} := \widehat{r} - (\widehat{r}\cdot\widehat{v}_{s})\widehat{v}_{s},
\tag{M16}
\end{equation}
switching to another reference (e.g. $\widehat{y}$) if $\|\vec{u}_{1}\|$ is below a small tolerance. Define the orthonormal pair in $\mathcal{P}$:
\begin{equation}
\widehat{u}_{1}^{(0)} = \frac{\vec{u}_{1}}{\|\vec{u}_{1}\|},
\qquad
\widehat{u}_{2}^{(0)} = \widehat{v}_{s}\times\widehat{u}_{1}^{(0)},
\tag{M17}
\end{equation}
and then apply the in-plane ``symmetric-gauge'' rotation by $\theta=\pi/4$ to obtain the pair $\widehat{u}_{1},\widehat{u}_{2}$:
\begin{equation}
\begin{bmatrix}
\widehat{u}_{1} & \widehat{u}_{2}
\end{bmatrix}
=
\begin{bmatrix}
\widehat{u}_{1}^{(0)} & \widehat{u}_{2}^{(0)}
\end{bmatrix}
\overleftrightarrow{R}_{s}\left(\frac{\pi}{4}\right),
\qquad
\overleftrightarrow{R}_{s}(\theta)
=
\begin{bmatrix}
\cos\theta & -\sin\theta\\
\sin\theta & \cos\theta
\end{bmatrix}
\in SO(2,\mathbb{R}).
\tag{M18}
\end{equation}
Writing
\begin{equation}
\overleftrightarrow{R}_{ij}
=
\begin{bmatrix}
\widehat{u}_{1} & \widehat{u}_{2} & \widehat{v}_{s}
\end{bmatrix}
\in SO(3,\mathbb{R}),
\qquad
\overleftrightarrow{V}^{(ij)}
= \overleftrightarrow{R}_{ij}^{T}\overleftrightarrow{V}\overleftrightarrow{R}_{ij},
\tag{M19}
\end{equation}
the tensor assumes the block form (up to numerical tolerance):
\begin{equation}
\overleftrightarrow{V}^{(ij)}
=
\begin{bmatrix}
\overleftrightarrow{A}^{(ij)} & 0_{2\times1}\\
0_{1\times2} & \lambda_s
\end{bmatrix},
\qquad
\overleftrightarrow{A}^{(ij)}
=
\begin{bmatrix}
\widehat{u}_{1}^{T}\overleftrightarrow{V}\widehat{u}_{1} & \widehat{u}_{1}^{T}\overleftrightarrow{V}\widehat{u}_{2}\\
\widehat{u}_{2}^{T}\overleftrightarrow{V}\widehat{u}_{1} & \widehat{u}_{2}^{T}\overleftrightarrow{V}\widehat{u}_{2}
\end{bmatrix}.
\tag{M20}
\end{equation}
We note that $\overleftrightarrow{A}^{(ij)}$ is real, symmetric and contains the two participating eigenvalues. The choices $ij\in\{12,13,23\}$ correspond to three conventions for which the eigenvalue line is treated as a spectator (i.e. which complementary $s$ is selected).

Along the discretized loop, we fix signs by continuity, e.g. enforce for iterative $n=\{1,2,3,\ldots\}$, $\widehat{v}_{s}^{(n)}\cdot\widehat{v}_{s}^{(n-1)}>0$ and $\widehat{u}_{1}^{(n)}\cdot\widehat{u}_{1}^{(n-1)}>0$ to avoid artificial jumps in the winding angle.

\subsubsection*{(C2) Two-level Hamiltonian and the degeneracy condition}

Defining $A_{ab}=\widehat{u}_{a}^{T}\overleftrightarrow{V}\widehat{u}_{b}$ for $a,b\in\{1,2\}$, we can decompose $\overleftrightarrow{A}^{(ij)}$ into its trace and traceless parts:
\begin{equation}
t := \frac{1}{2}(A_{11}+A_{22}),
\qquad
d_{z} := \frac{1}{2}(A_{11}-A_{22}),
\qquad
d_{x} := A_{12},
\tag{M21}
\end{equation}
so that:
\begin{equation}
\overleftrightarrow{A}^{(ij)}
= t\overleftrightarrow{Id}+\overleftrightarrow{H},
\qquad
\overleftrightarrow{H}
=
\begin{bmatrix}
d_z & d_x\\
d_x & -d_z
\end{bmatrix}
= d_z\overleftrightarrow{\sigma_z}+d_x\overleftrightarrow{\sigma_x},
\tag{M22}
\end{equation}
where
\begin{equation*}
\overleftrightarrow{\sigma_x}=
\begin{bmatrix}
0&1\\
1&0
\end{bmatrix},
\qquad
\overleftrightarrow{\sigma_z}=
\begin{bmatrix}
1&0\\
0&-1
\end{bmatrix}
\end{equation*}
are the Pauli matrices. 

The imaginary Pauli matrix $\overleftrightarrow{\sigma_y}$ cannot appear because $\overleftrightarrow{A}^{(ij)}$ is real symmetric, and $\{\overleftrightarrow{Id},\overleftrightarrow{\sigma_x},\overleftrightarrow{\sigma_z}\}$ spans the space of real symmetric $2\times2$ matrices. This real-symmetric structure is associated with time-reversal symmetry $\mathcal{T}=K$ and $\mathcal{T}^{2}=+1$, as in class AI of the Altland--Zirnbauer classification \cite{Altland1997AZ,Chiu2016RMP}.

The traceless part in Eq.~(M22) is algebraically analogous to the two-band form of chiral models such as the Su--Schrieffer--Heeger model \cite{Su1979SSH,Su1980SSH}, but here it describes a real symmetric tensor subblock rather than a Bloch Hamiltonian. The integer $w$ defined in Methods (C3) depends on the orientation of the in-plane frame fixed in Eq.~(M18), whereas only its parity enters the eigenframe holonomy.

The splitting between the participating eigenvalues is:
\begin{equation}
\Delta\lambda_{ij}=2\sqrt{d_x^2+d_z^2},
\tag{M23}
\end{equation}
and the double degeneracy occurs if and only if $d_x=d_z=0$, i.e. $\overleftrightarrow{A}^{(ij)}\propto\overleftrightarrow{Id}$.

In computations, we treat points with $d_x^2+d_z^2<\varepsilon^2$, with a small $\varepsilon$ up to machine precision, as numerically indistinguishable from the degeneracy and exclude them from winding evaluation.

\subsubsection*{(C3) $\mathbb{Z}_{2}$ holonomy from planar winding}

Along a closed loop $C$, the pair $(d_z,d_x)$ traces a closed curve in the plane. Define
\begin{equation}
\phi := \operatorname{atan2}(d_x,d_z),
\tag{M24}
\end{equation}
such that $(d_z,d_x)=r(\cos\phi,\sin\phi)$, with $r=\sqrt{d_x^2+d_z^2}$. The winding number of $C$ about the degeneracy point $(d_z,d_x)=(0,0)$ is:
\begin{equation}
w = \frac{1}{2\pi}\oint_{C} d\phi \in \mathbb{Z}.
\tag{M25}
\end{equation}

For a real two-level system, continuous transport of a normalized real eigenvector of $\overleftrightarrow{H}$ around $C$ returns itself up to a sign:
\begin{equation}
\widehat{u}(2\pi)=(-1)^w\widehat{u}(0),
\tag{M26}
\end{equation}
so the holonomy (and hence the topological charge) is the first Stiefel--Whitney invariant~\cite{von_Neumann_J_1929,Herzberg_G_1963,Ahn_J_2019}:
\begin{equation}
w_1 := w \mod 2 \in \mathbb{Z}_2.
\tag{M27}
\end{equation}
It follows that this invariant $w_1$ indicates whether a continuously transported arrow along the headless eigenline returns unchanged, $w_1=0$, or reversed, $w_1=1$. 

For an elementary degeneracy ($w=\pm1$), the signed half-winding $w/2$ is the half-integer index familiar from tensor-field topology~\cite{Berry_Hannay_1977,Delmarcelle_T_1994}. We do not report it here: the transported eigenline records only $(-1)^{w}$ through Eq.~(M26), so loops of winding $+1$ and $-1$ are indistinguishable in the eigenframe, and the sign of $w$ in our calculations depends (in any case) on the sign returned for the spectator eigenvector $\widehat{v}_{s}$, which fixes the handedness of the in-plane frame in Eq.~(M18). Hence, we retain the $\mathbb{Z}_{2}$ charge $w_{1}$.

For a discretized loop $\{(d_{z,n},d_{x,n})\}_{n=0}^{N}$, we compute $\phi_n=\operatorname{atan2}(d_{x,n},d_{z,n})$, unwrap $\phi_n$, and estimate:
\begin{equation}
w = \operatorname{round}\left(\frac{\phi_N-\phi_0}{2\pi}\right),
\qquad
w_1 = w \mod 2.
\tag{M28}
\end{equation}
Equivalently, we compute the incremental winding using complex ratios $z_n=d_{z,n}+i d_{x,n}$:
\begin{equation}
w = \frac{1}{2\pi}\sum_{n=1}^{N}\Delta\phi_n,
\qquad
\Delta\phi_n=\arg\left(\frac{z_n}{z_{n-1}}\right)\in(-\pi,\pi].
\tag{M29}
\end{equation}

\subsubsection*{(C4) Eigenvector transport and sign-flip detection}

To verify the $\mathbb{Z}_{2}$ classification obtained from the winding of $(d_z,d_x)$, we perform explicit eigenvector transport along the loop. For each sampled point $n$, we diagonalize $\overleftrightarrow{A}_{n}^{(ij)}$ and choose the normalized eigenvector $\widehat{c}_{n}\in\mathbb{R}^{2}$ corresponding to the tracked branch:
\begin{equation}
\widehat{c}_{n}=
\begin{pmatrix}
c_{n,1}\\
c_{n,2}
\end{pmatrix}.
\tag{M30}
\end{equation}
Because real eigenvectors are defined only up to an overall sign, we enforce a continuous choice by fixing the sign such that successive vectors have positive overlap:
\begin{equation}
\widehat{c}_{n}\cdot\widehat{c}_{n-1}>0,
\tag{M31}
\end{equation}
flipping $\widehat{c}_{n}\rightarrow-\widehat{c}_{n}$ when necessary.

The corresponding three-dimensional eigenvector is reconstructed as (cf. Eq. M18)
\begin{equation}
\psi_n
= c_{n,1}\widehat{u}_{1}^{(n)} + c_{n,2}\widehat{u}_{2}^{(n)}
\in \mathcal{P}^{(n)}\subset\mathbb{R}^{3}.
\tag{M32}
\end{equation}
A nontrivial holonomy is present if the continuously transported eigenvector returns with opposite sign, i.e. $\psi_N\cdot\psi_0<0$, and if the loop $C$ encircles but does not pass through the degeneracy where $d_x^2+d_z^2<\varepsilon^2$, with a noise tolerance $\varepsilon$.

The $\mathbb{Z}_{2}$ invariant is diagnosed both by the winding of the reduced two-level effective field and by direct continuous transport of the corresponding eigenframe; agreement of these two constructions confirms that the classification is intrinsic and not a consequence of a particular local gauge choice.

\subsection*{(D) Symmetry-adapted tensor basis for MgO}

MgO provides a symmetry-clean reference because the Mg site has octahedral symmetry $O_h$ and the unstrained EFG vanishes ($\overleftrightarrow{V}=\overleftrightarrow{0}$) by symmetry. In the small-deformation (linear-response) regime, the deviatoric (traceless) part of the applied stress/strain spans the same five-dimensional vector space $X=\operatorname{Sym}_{0}(3,\mathbb{R})$ as the EFG tensor, so the five independent deviatoric control components act as symmetry-adapted ``knobs'' for generating an arbitrary local EFG within $X$.

\subsubsection*{(D1) Frobenius-orthonormal basis for $X$}

We equip the space of real symmetric tensors with the Frobenius inner product
\begin{equation}
\langle \overleftrightarrow{A},\overleftrightarrow{B}\rangle
= \operatorname{Tr}(\overleftrightarrow{A}^{T}\overleftrightarrow{B})
= \sum_{i=1}^{3}\sum_{j=1}^{3}A_{ij}B_{ij},
\qquad
\overleftrightarrow{A},\overleftrightarrow{B}\in\operatorname{Sym}(3,\mathbb{R}).
\tag{M33}
\end{equation}
and decompose $\operatorname{Sym}(3,\mathbb{R})=\mathbb{R}\overleftrightarrow{Id}\oplus X$, i.e., into the direct sum of its isotropic and its traceless part.

A convenient Frobenius-orthonormal basis of $X$ is
\begin{align}
\widehat{\varepsilon}_{E_{g,\mathrm{octa}\,(x^2-y^2)}}
&=\frac{1}{\sqrt{2}}
\begin{bmatrix}
1&0&0\\
0&-1&0\\
0&0&0
\end{bmatrix},
&
\widehat{\varepsilon}_{E_{g,\mathrm{tetra}\,(3z^2-r^2)}}
&=\frac{1}{\sqrt{6}}
\begin{bmatrix}
1&0&0\\
0&1&0\\
0&0&-2
\end{bmatrix},
\notag\\[2mm]
\widehat{\varepsilon}_{T_{2g,xy}}
&=\frac{1}{\sqrt{2}}
\begin{bmatrix}
0&1&0\\
1&0&0\\
0&0&0
\end{bmatrix},
&
\widehat{\varepsilon}_{T_{2g,xz}}
&=\frac{1}{\sqrt{2}}
\begin{bmatrix}
0&0&1\\
0&0&0\\
1&0&0
\end{bmatrix},
\notag\\[2mm]
\widehat{\varepsilon}_{T_{2g,yz}}
&=\frac{1}{\sqrt{2}}
\begin{bmatrix}
0&0&0\\
0&0&1\\
0&1&0
\end{bmatrix}.
\tag{M34a}
\end{align}
These five matrices satisfy $\langle \widehat{\varepsilon}_{\alpha},\widehat{\varepsilon}_{\beta}\rangle=\delta_{\alpha,\beta}, \alpha \in\{E_{g,\mathrm{octa}\,(x^2-y^2)},E_{g,\mathrm{tetra}\,(3z^2-r^2)},T_{2g,xy},T_{2g,xz},T_{2g,yz}\}$ and span $X$. For completion, the missing sixth symmetric basis element:
\begin{equation}
\widehat{\varepsilon}_{A_{1g,r^2}}
=\frac{1}{\sqrt{3}}
\begin{bmatrix}
1&0&0\\
0&1&0\\
0&0&1
\end{bmatrix}
\tag{M34b}
\end{equation}
is the isotropic component $\overleftrightarrow{Id}/\sqrt{3}$, orthogonal to $X$.

Given any $\overleftrightarrow{S}\in X$, its coordinates in this basis are obtained by projection:
\begin{equation}
s_{\alpha}=\langle \overleftrightarrow{S},\overleftrightarrow{B}_{\alpha}\rangle,
\qquad
\overleftrightarrow{S}=\sum_{\alpha=1}^{5}s_{\alpha}\overleftrightarrow{B}_{\alpha},
\tag{M35}
\end{equation}
with
\begin{equation*}
\overleftrightarrow{B}_{\alpha}\in
\left\{
\widehat{\varepsilon}_{E_{g,\mathrm{octa}\,(x^2-y^2)}},
\widehat{\varepsilon}_{E_{g,\mathrm{tetra}\,(3z^2-r^2)}},
\widehat{\varepsilon}_{T_{2g,xy}},
\widehat{\varepsilon}_{T_{2g,xz}},
\widehat{\varepsilon}_{T_{2g,yz}}
\right\}.
\end{equation*}

\subsubsection*{(D2) $O_h$ decomposition into $E_g \oplus T_{2g}$}

Under the natural $O_h$ action:
\begin{equation}
T(g):X\rightarrow X,\quad T(g)(\overleftrightarrow{Y})=\overleftrightarrow{R}(g)^{T}\overleftrightarrow{Y}\overleftrightarrow{R}(g),
\qquad g\in O_h,
\tag{M36}
\end{equation}
the five-dimensional space $X$ decomposes into irreducible subspaces
\begin{equation}
X\cong E_g\oplus T_{2g},
\tag{M37}
\end{equation}
where $\{\widehat{\varepsilon}_{E_{g,\mathrm{octa}\,(x^2-y^2)}},\widehat{\varepsilon}_{E_{g,\mathrm{tetra}\,(3z^2-r^2)}}\}$ span the $E_g$ subspace and $\{\widehat{\varepsilon}_{T_{2g,xy}},\widehat{\varepsilon}_{T_{2g,xz}},\widehat{\varepsilon}_{T_{2g,yz}}\}$ span the $T_{2g}$ subspace.

\subsubsection*{(D3) Strain coordinates and linear EFG response}

We parameterize a small homogeneous strain by the infinitesimal tensor $\overleftrightarrow{\varepsilon}$. Its deviatoric part,
\begin{equation}
\widetilde{\varepsilon}
= \overleftrightarrow{\varepsilon}-\frac{\operatorname{Tr}(\overleftrightarrow{\varepsilon})}{3}\overleftrightarrow{Id}
\in X,
\tag{M38}
\end{equation}
is expanded in the $O_h$-adapted basis:
\begin{equation}
\widetilde{\varepsilon}
= q_{E_{\mathrm{octa}}}\widehat{\varepsilon}_{E_{g,\mathrm{octa}\,(x^2-y^2)}}
+ q_{E_{\mathrm{tetra}}}\widehat{\varepsilon}_{E_{g,\mathrm{tetra}\,(3z^2-r^2)}}
+ q_{xy}\widehat{\varepsilon}_{T_{2g,xy}}
+ q_{xz}\widehat{\varepsilon}_{T_{2g,xz}}
+ q_{yz}\widehat{\varepsilon}_{T_{2g,yz}},
\tag{M39}
\end{equation}
with
\begin{equation*}
q_{\alpha}=\langle\widetilde{\varepsilon},\overleftrightarrow{B}_{\alpha}\rangle.
\end{equation*}
Because the EFG is traceless and symmetric, the isotropic $A_{1g}$ strain component (proportional to $\overleftrightarrow{Id}$) does not contribute to $\overleftrightarrow{V}$ at linear order in MgO.

In the linear-response regime we model the strain-induced EFG as a linear map $\overleftrightarrow{L}:X\rightarrow X$,
\begin{equation}
\overleftrightarrow{V}=\overleftrightarrow{L}(\widetilde{\varepsilon}),
\tag{M40}
\end{equation}
and octahedral symmetry constrains $L$ to act independently on the $E_g$ and $T_{2g}$, acting as a single scalar on each, by Schur's lemma \cite{Dresselhaus2008}. Operationally, we therefore treat $E_g$-type controls as primarily affecting the diagonal components of $\overleftrightarrow{V}$, and $T_{2g}$-type controls as primarily affecting its off-diagonal components.

\subsubsection*{(D4) Degeneracy loci in the
$E_g^{\mathrm{tet}}-T_{2g}^{xy}$ plane}

Consider the two-mode strain family used in Fig.~\ref{fig:MgO}, spanned by the
$E_g^{\mathrm{tet}}$ and $T_{2g}^{xy}$ directions defined in Eq.~(M34a)
and sampled according to the procedure in Methods E3. Here,
$E_g^{\mathrm{tet}}$ denotes the
$E_{g,\mathrm{tetra}\,(3z^2-r^2)}$ basis direction. We denote the
computational amplitudes of these two modes by $t$ and $s$, respectively.
Any fixed normalization difference between these amplitudes and the
Frobenius coordinates introduced in Eq.~(M39) is absorbed into the
response functions below.

Throughout this control plane, the residual crystal symmetries constrain
the EFG tensor to
\begin{equation}
\overleftrightarrow{V}(t,s)=
\begin{pmatrix}
a(t,s) & b(t,s) & 0\\
b(t,s) & a(t,s) & 0\\
0 & 0 & -2a(t,s)
\end{pmatrix},
\tag{M41a}
\end{equation}
where $a$ and $b$ include both linear and nonlinear strain responses. The residual symmetries preserve three mutually orthogonal eigenaxes. 

Choosing normalized vector representatives:
\begin{equation}
\begin{aligned}
\boldsymbol{v}_{+}
&=
\frac{1}{\sqrt{2}}
\begin{pmatrix}
1\\
1\\
0
\end{pmatrix},
&\qquad
\lambda_{+}&=a+b,\\[1mm]
\boldsymbol{v}_{-}
&=
\frac{1}{\sqrt{2}}
\begin{pmatrix}
1\\
-1\\
0
\end{pmatrix},
&
\lambda_{-}&=a-b,\\[1mm]
\boldsymbol{v}_{z}
&=
\begin{pmatrix}
0\\
0\\
1
\end{pmatrix},
&
\lambda_{z}&=-2a.
\end{aligned}
\tag{M41b}
\label{eq:M41b}
\end{equation}
The vectors in Eq.~\eqref{eq:M41b} are representatives of headless principal axes, so their overall signs have no physical significance.

Away from $\overleftrightarrow{V}=\overleftrightarrow{0}$, the axial condition $\eta=0$ is reached when any two of the three symmetry-labelled principal values coincide:
\begin{equation*}
\lambda_{+}=\lambda_{-}
\iff b=0,
\qquad
\lambda_{+}=\lambda_{z}
\iff b=-3a,
\qquad
\lambda_{-}=\lambda_{z}
\iff b=3a.
\end{equation*}
At the cubic point $a=b=0$, all three principal values vanish and $\eta$ is undefined.

Along $s=0$, tetragonal symmetry is restored and enforces $b(t,0)=0$, producing the symmetry-pinned vertical degeneracy branch. The two remaining branches arise when an in-plane eigenvalue equals the $z$-sector eigenvalue. The residual symmetries forbid $xz$ and $yz$ couplings, so these sectors cannot mix and their crossings remain exact. However, symmetry does not prescribe the strain values at which $b=\pm3a$ is satisfied. Therefore, these material-dependent loci can shift or bend through nonlinear response and are described as symmetry-allowed, unpinned crossings. A perturbation that permits $xz$ or $yz$ coupling would convert them into avoided crossings. 

In this control plane, each degeneracy curve was determined by locating the strain values at which the corresponding signed eigenvalue difference in Eq.~\eqref{eq:M41b} vanishes. As the symmetry-labelled eigenvalue branches vary smoothly, their difference also varies smoothly and is precisely zero where the two principal values coincide.

\subsection*{(E) First-principles EFG control trajectories}

\subsubsection*{(E1) Pseudopotentials}

PAW datasets (VASP POTCAR) were
\begin{enumerate}
\renewcommand{\labelenumi}{\arabic{enumi})}
\item PAW\_PBE Ti\_sv and PAW\_PBE O for TiO$_2$,
\item PAW\_PBE Sn\_sv\_GW and PAW\_PBE O for SnO$_2$, and
\item PAW\_PBE Mg\_sv and PAW\_PBE O for MgO.
\end{enumerate}
As the EFGs are sensitive to near-nucleus charge density~\cite{Petrilli_H_M_1998}, semicore (``sv'') datasets were used for Ti, Sn, and Mg; the ``GW'' tag in Sn\_sv\_GW is part of the dataset name and does not imply GW calculations.

\subsubsection*{(E2) Geometry Optimization}

Geometry optimization of both rutile TiO$_2$ and SnO$_2$ was performed within density-functional theory using the PBE generalized-gradient approximation (GGA = PE) and a plane-wave cutoff of 600 eV (ENCUT = 600). Brillouin-zone integrations employed the tetrahedron method with Blöchl corrections (ISMEAR = $-5$, SIGMA = 0.01) on a $\Gamma$-centered $15\times15\times15$ Monkhorst--Pack grid (KPOINTS: ``Gamma'', 15 15 15). Electronic self-consistency was converged to $10^{-10}$ eV (EDIFF = 1E-10) with a maximum of 200 electronic steps (NELM = 200, NELMIN = 6) using the standard mixing/diagonalization scheme (ALGO = Normal). Structural relaxation was carried out using the conjugate-gradient algorithm (IBRION = 2) with relaxation of ionic positions, cell shape, and volume (ISIF = 3) for up to 200 ionic steps (NSW = 200), terminating when forces satisfied $10^{-6}$ eV/\AA{} (EDIFFG = $-1$E-6). Electric-field gradients were output during the calculation (LEFG = .TRUE.).

MgO calculations were performed in VASP using the PAW method with PREC = Accurate and a plane-wave cutoff of 800 eV (ENCUT = 800). Symmetry was enabled (ISYM = 2) with a tight symmetry tolerance (SYMPREC = 1E-8) to enforce the cubic $O_h$ structure in the unstrained reference. Electric-field gradients were computed (LEFG = .TRUE.) with nonspherical contributions included (LASPH = .TRUE.), accurate real-space projection (LREAL = .FALSE.), and an additional support grid (ADDGRID = .TRUE.). Brillouin-zone sampling used a Monkhorst--Pack $7\times7\times7$ mesh. Electronic self-consistency employed Gaussian smearing (ISMEAR = 0, SIGMA = 0.01), ALGO = Normal, EDIFF = 1E-11, and NELM = 200. Structural relaxation (where performed) used the conjugate-gradient algorithm (IBRION = 2) allowing relaxation of ions, cell shape, and volume (ISIF = 3) for up to 200 ionic steps (NSW = 200), with a force convergence threshold of $10^{-6}$ eV/\AA{} (EDIFFG = $-1$E-6) and POTIM = 0.2. For selected runs, the FFT grid was fixed to NGXF = NGYF = NGZF = 192.

\subsubsection*{(E3) Strain Calculations}

Strain control trajectories in both rutile TiO$_2$ and SnO$_2$ were generated by applying the in-plane symmetry-adapted strain tensor
\begin{equation}
\overleftrightarrow{\varepsilon}(Q_{B_{1g}},Q_{B_{2g}})
=
Q_{B_{1g}}\,\widehat{\varepsilon}_{B_{1g}}
+
Q_{B_{2g}}\,\widehat{\varepsilon}_{B_{2g}}
=
\begin{pmatrix}
Q_{B_{1g}} & Q_{B_{2g}} & 0\\
Q_{B_{2g}} & -Q_{B_{1g}} & 0\\
0 & 0 & 0
\end{pmatrix},
\tag{M42}
\end{equation}
corresponding to B$_{1g}$-type orthorhombic strain ($\varepsilon_{xx}=-\varepsilon_{yy}$) and B$_{2g}$-type shear ($\varepsilon_{xy}$), with the symmetry-adapted basis elements $\widehat{\varepsilon}_{B_{1g}}$ and $\widehat{\varepsilon}_{B_{2g}}$ as defined in the main text. Here, $Q_{B_{1g}}$ and $Q_{B_{2g}}$ are dimensionless strain amplitudes (e.g. $Q_{B_{1g}}=0.01$ corresponds to a $\pm1\%$ in-plane deviatoric strain). Lattice vectors $\vec{A}$ were updated via the deformation gradient $\overleftrightarrow{F}=\overleftrightarrow{Id}+\overleftrightarrow{\varepsilon}$, i.e. $\vec{A}^{\,\prime}=\overleftrightarrow{F}\vec{A}$. Fractional atomic coordinates were kept fixed under the affine deformation and internal coordinates were subsequently relaxed at fixed strained cell (IBRION = 2, ISIF = 2, NSW = 200, POTIM = 0.2). 

Self-consistent calculations were performed in VASP using the PAW method within PBE (GGA = PE) with PREC = Accurate, ENCUT = 600 eV, and a $\Gamma$-centered $15\times15\times15$ k-point mesh; Brillouin-zone integrations used the tetrahedron method with Blöchl corrections (ISMEAR = $-5$, SIGMA = 0.01), with ALGO = Normal, NELM = 200, and electronic convergence EDIFF = $10^{-10}$ eV. Relaxations were terminated at EDIFFG = $-10^{-5}$ eV/\AA{}. Electric-field gradients were extracted from the converged calculations (LEFG = .TRUE., LREAL = .FALSE.) for subsequent invariant/PAS analysis (Methods A--C).

Strain control trajectories in rocksalt MgO were generated using symmetry-adapted $O_h$ strain modes (Methods D), by forming a small-strain tensor $\overleftrightarrow\varepsilon(q_{1},q_{2})=q_{1}\overleftrightarrow\varepsilon^{(1)}+q_{2}\overleftrightarrow\varepsilon^{(2)}$ (e.g. $T_{2g,xy}:\varepsilon_{xy}=q_{1}$, $T_{2g,xz}:\varepsilon_{xz}=q_{2}$, $E_g$ diagonal modes, and $A_{1g}$ hydrostatic). The mode amplitudes $q_{1,2}$ are dimensionless strains (reported in percent; e.g. $q=0.03$ corresponds to 3\%). 

The deformation gradient $\overleftrightarrow{F}=\overleftrightarrow{Id}+\overleftrightarrow{\varepsilon}$ was applied to the primitive lattice vectors, fractional coordinates were kept fixed under the affine deformation, and internal coordinates were subsequently relaxed at fixed strained cell (IBRION = 2, ISIF = 2, NSW = 200, EDIFFG = $-10^{-6}$ eV/\AA{}). Self-consistent calculations used PREC = Accurate, ENCUT = 800 eV, a Monkhorst--Pack $7\times7\times7$ k-point mesh, Gaussian smearing (ISMEAR = 0, SIGMA = 0.01), ALGO = Normal, EDIFF = $10^{-10}$ eV, and NELM = 200; EFGs were extracted with LEFG = .TRUE., using LASPH = .TRUE., ADDGRID = .TRUE., and LREAL = .FALSE. Symmetry was enabled with ISYM = 2 and SYMPREC = $10^{-8}$; for selected runs the FFT grid was fixed to NGXF = NGYF = NGZF = 192.

\subsubsection*{(E4) Verification of the TiO$_2$ degeneracy}
 
In the $Q_{B_{1g}}\,$-$\,Q_{B_{2g}}$ strain plane, symmetry preserves the $xy\oplus z$ block form
\begin{equation}
\overleftrightarrow{V}(\vec{Q})
=
\begin{pmatrix}
t(\vec{Q})+d_z(\vec{Q}) & d_x(\vec{Q}) & 0\\
d_x(\vec{Q}) & t(\vec{Q})-d_z(\vec{Q}) & 0\\
0 & 0 & -2t(\vec{Q})
\end{pmatrix},
\qquad
\vec{Q}=(Q_{B_{1g}},Q_{B_{2g}}),
\tag{M43}
\end{equation}
where
\begin{equation}
t=\frac{V_{xx}+V_{yy}}{2},
\qquad
d_z=\frac{V_{xx}-V_{yy}}{2},
\qquad
d_x=V_{xy}.
\tag{M44}
\end{equation}
The two in-plane principal values are
\begin{equation}
\lambda_{\pm}
=
t\pm\sqrt{d_z^2+d_x^2},
\tag{M45}
\end{equation}
and coincide when $V_{xx}=V_{yy}$ and $V_{xy}=0$.

The EFG components, including $t$, $d_z$, and $d_x$, are reported in $\mathrm{V\,\AA^{-2}}$. Since $Q_{B_{1g}}$ and $Q_{B_{2g}}$ are dimensionless strain amplitudes, the entries of the Jacobian $J$ are reported in $\mathrm{V\,\AA^{-2}}$ per unit strain.
 
The Cartesian EFG components were fitted over the sampled strain range using
\begin{equation}
\begin{split}
V_{ij}(\vec{Q})={}&
a_{ij}^{(0)}
+a_{ij}^{(1)}Q_{B_{1g}}
+a_{ij}^{(2)}Q_{B_{2g}}
+a_{ij}^{(11)}Q_{B_{1g}}^2\\
&+a_{ij}^{(22)}Q_{B_{2g}}^2
+a_{ij}^{(12)}Q_{B_{1g}}Q_{B_{2g}}.
\end{split}
\tag{M46}
\end{equation}
The fits yield a common zero within the sampled strain range at
\begin{equation}
\vec{Q}^{*}
=
\left(
-1.21\times10^{-8},
-1.20387\times10^{-2}
\right)
\simeq
\left(0,-1.204\%\right).
\tag{M47}
\end{equation}
At this point, $t(\vec{Q}^{*})=6.88538\neq0$, confirming a double rather than triple degeneracy.
 
The local Jacobian of $(d_z,d_x)$ with respect to the two strain amplitudes is
\begin{equation}
J(\vec{Q}^{*})
=
\left.
\begin{pmatrix}
\partial_{Q_{B_{1g}}}d_z &
\partial_{Q_{B_{2g}}}d_z\\
\partial_{Q_{B_{1g}}}d_x &
\partial_{Q_{B_{2g}}}d_x
\end{pmatrix}
\right|_{\vec{Q}=\vec{Q}^{*}}
=
\begin{pmatrix}
-210.4066 & -6.85\times10^{-5}\\
-1.60\times10^{-3} & 644.1869
\end{pmatrix}.
\tag{M48}
\end{equation}
Since
\begin{equation}
\det J(\vec{Q}^{*})
=
-1.35541\times10^{5}\neq0,
\tag{M49}
\end{equation}
the common zero is isolated and the $Q_{B_{1g}}\,$-$\,Q_{B_{2g}}$ strain plane intersects the degeneracy stratum $\Sigma_0$ transversely.




\end{appendices}


\newpage 

\bibliography{sn-bibliography}

\end{document}